\documentclass[fleqn,usenatbib,useAMS]{mnras}

\usepackage{siunitx} 

\usepackage{graphicx}	
\usepackage{amsmath}	
\usepackage{amssymb}	
\usepackage{multicol}        
\usepackage{bm}		
\usepackage{pdflscape}	

\newcommand{\kms}{\,km\,s$^{-1}$} 

\usepackage[T1]{fontenc}
\usepackage{ae,aecompl}

\usepackage{newtxtext,newtxmath}

\title[Gas \& dust production in evolved planetary systems]{The dust never settles: collisional production of gas and dust in evolved planetary systems}

\author
[A. Swan et al.]
{Andrew Swan$^{1}$\thanks{E-mail: \href{mailto:a.swan.17@ucl.ac.uk}{a.swan.17@ucl.ac.uk}},
Jay Farihi$^{1}$,
Thomas G. Wilson$^{1,2}$,
and Steven G. Parsons$^{3}$
\\
$^{1}$Department of Physics~\& Astronomy, University College London, Gower Street, London~WC1E~6BT, UK\\
$^{2}$School of Physics~\& Astronomy, University of St~Andrews, North Haugh, St~Andrews, Fife~KY16~9SS, UK\\
$^{3}$Department of Physics and Astronomy, University of Sheffield, Sheffield~S3~7RH, UK\\
}

\date{Accepted XXX. Received YYY; in original form ZZZ}

\pubyear{2020}

\begin{document}
\label{firstpage}
\pagerange{\pageref{firstpage}--\pageref{lastpage}}
\maketitle

\begin{abstract}
Multi-epoch infrared photometry from \textit{Spitzer} is used to monitor circumstellar discs at white dwarfs, which are consistent with disrupted minor planets whose debris is accreted and chemically reflected by their host stars. Widespread infrared variability is found across the population of 37 stars with two or more epochs. Larger flux changes occur on longer time-scales, reaching several tens of per~cent over baselines of a few years. The canonical model of a geometrically thin, optically thick disc is thus insufficient, as it cannot give rise to the observed behaviour. Optically thin dust best accounts for the variability, where collisions drive dust production and destruction. Notably, the highest infrared variations are seen in systems that show \ion{Ca}{ii}~emission, supporting planetesimal collisions for all known debris discs, with the most energetic occurring in those with detected gaseous debris. The sample includes the only polluted white dwarf with a circumbinary disc, where the signal of the day--night cycle of its irradiated substellar companion appears diluted by dust emission.
\end{abstract}




\section{Introduction}
\label{sectionIntroduction}

The first infrared excess detected at a white dwarf was originally attributed to a brown dwarf companion \citep{Zuckerman1987}, but circumstellar dust soon became the favoured hypothesis \citep{Wickramasinghe1988, Tokunaga1990}. Almost two decades passed before a second white dwarf dust disc was identified \citep{Becklin2005, Kilic2005}. The \textit{Spitzer Space Telescope} \citep{Werner2004Spitzer} enabled sensitive searches for infrared excesses among large samples of white dwarfs, with less than three per~cent showing detectable dust \citep{Jura2007Spitzer11stars, Mullally2007, Farihi2009Spitzer, Barber2012, Bergfors2014, Rocchetto2015, Wilson2019}. Brown dwarf companions to white dwarfs are rarer still \citep{Farihi2005, Girven2011, Debes2011}, and only one is known where the system also hosts dust (SDSS\,J155720.77+091624.6, hereafter SDSS\,1557; \citealt{Farihi2017SDSS1557}). Over 40~dusty white dwarfs are known at present, and a similar number of new candidates identified from \textit{Gaia} and \textit{WISE} await follow-up \citep{Rebassa-Mansergas2019}.

Dust is proposed to arise from tidal disruption of minor planets, and transiting clouds of debris have been observed near the stellar Roche limit that are consistent with that model \citep{Jura2003, Vanderburg2015}. Photospheric metal pollution always accompanies circumstellar debris, providing a compelling link to planetary systems, as the accreted material almost universally has elemental abundances consistent with rock \citep{Gansicke2012, Jura2014review, Hollands2018analysis, Xu2019composition}. Some dusty systems display emission and absorption by gaseous debris coincident with dust \citep{Brinkworth2009, Melis2010, Cauley2018}, and targeted searches for \ion{Ca}{ii} emission confirm they form a distinct subset \citep{Dennihy2018, Manser2020}. There is also spectroscopic evidence of a planetesimal on a two-hour orbit in one such system \citep{Manser2019}. While planetary bodies have yet to be directly detected around white dwarfs, these evolved systems nevertheless provide the most direct access to the bulk composition of exoplanets.

Many studies of dusty white dwarfs employ a model where a circular, geometrically thin, optically thick annulus resides interior to the stellar Roche limit and exterior to where dust rapidly sublimates \citep{Jura2003}. This is referred to here as the \textit{canonical disc}. It has been found to fit most observed infrared excesses well (e.g. \mbox{\citealt{Rocchetto2015}}), but there is a growing body of evidence that this picture is incomplete (see also the discussion in \citealt{Nixon2020}). The infrared brightness of systems that display \ion{Ca}{ii} emission from high-inclination gaseous debris is surprising, as the canonical disc is brightest when seen face-on, suggesting an additional component to the infrared emission \citep{Dennihy2017}. The presence of \mbox{optically} thin dust is required to fit the excesses at some systems \citep{Jura2007GD362, Farihi2017SDSS1557, Farihi2018GD56}, and has been proposed to account for the absence of infrared excesses at many polluted stars \citep{Bonsor2017}.

Periodic near-infrared variation was found at the prototypical dusty white dwarf, 2326$-$049 (G29-38)\footnote{By convention, stars are identified throughout by their WD designation (their abbreviated B1950.0 coordinates)}, a pulsating ZZ\,Ceti-type star \citep{Patterson1991}. However, such variability received little attention until the infrared flux at 0956$-$017 dropped by 35~per~cent within a year \citep{Xu2014drop}. Further examples were uncovered from archival \textit{Spitzer} data \citep{Farihi2018GD56, Xu2018IRvariability}, and a decade of observations from the \textit{WISE} satellite revealed that variation at {3.5\,\micron} is commonplace among the general population of dusty white dwarfs \citep{Swan2019}. In contrast, variation appears diminished at shorter wavelengths: a ground-based survey observed no significant changes in \textit{K}-band brightness on time-scales up to three years \citep{Rogers2020}. The most spectacular event to date is the outburst at 0145+234, where the infrared light curve shows unprecedented brightening and colour changes \citep{Wang2019_0145}.

Several models predict or hypothesise conditions conducive to infrared variation in white dwarf debris discs. Flux changes can result from a collisional cascade that modifies the population of micron-sized particles that dominate infrared emission \citep{Wyatt2008}. Simulations exploring that process in the context of metal accretion find that the infrared flux can rise and fall stochastically \citep{Kenyon2017collisions, Kenyon2017gas}. Another study finds that a canonical disc can experience runaway accretion that rapidly removes circumstellar material, reducing the infrared emission \citep{Rafikov2011runaway}. Simulations show that complete tidal disruption of asteroids can occur within a few orbital periods, distributing clumpy material across a range of orbits (\citealt{Malamud2020A, Malamud2020B}; see also \citealt{Coughlin2015}), which can give rise to infrared variability \citep{Nixon2020}. However, the subsequent evolution of such material remains to be established. Poynting--Robertson~(PR) drag will shrink the orbits, but collisions will come to dominate before the material approaches a canonical disc configuration \citep{Farihi2008}.

Numerous physical processes were considered as potential causes of the dimming and brightening episodes at 0408$-$041, including collisions, sublimation and condensation, PR~drag, recycling of material into and out of small bodies outside the Roche radius, interaction between material on intersecting orbits, and disc warps induced by an external perturber \citep{Farihi2018GD56}. All of these may operate in white dwarf systems, but collisions were surmised to be likely drivers of infrared variation at 0408$-$041.

Infrared variation has been established as a characteristic of dusty white dwarfs. However, few examples have been examined in detail, and the physical processes responsible remain poorly constrained. That most dusty systems vary is an inference drawn from comparison against a control sample at the population level, where the data were insufficient for detailed analysis of individual targets \citep{Swan2019}. Better characterisation of flux changes at each star is required to identify trends within the population, while establishing the relevant time-scales will help to identify the mechanisms driving variation. Data continues to flow from \textit{WISE}, but it is limited by low sensitivity, a fixed six-monthly cadence, and source confusion \citep{Dennihy2020}. On the other hand, \textit{Spitzer} observations are capable of higher photometric precision, can access shorter time-scales, and suffer less contamination from background sources at greater depth. This paper reports on the flux changes at dusty white dwarfs observed with \textit{Spitzer}, using archival data supplemented with a photometric snapshot of the population. Observations and data analysis are described in Section~\ref{sectionObservations}, and results are presented in Section~\ref{sectionResults}, including a light curve of SDSS\,1557. The main results are summarised and discussed in Section~\ref{sectionDiscussion}.

\section{Observations and photometry}
\label{sectionObservations}

The 44~targets in program~13216 were observed between 2017~November and 2018~May, using the Infrared Array Camera (IRAC; \citealt{Fazio2004}) on \textit{Spitzer}. The only criterion for inclusion in the program was that a star should have an infrared excess consistent with dusty debris. Data were acquired in both the 3.6 and {4.5\,\micron} channels, employing the medium cycling dither pattern. Frame times were either 12 or 30\,s, and 10 or 20 exposures were taken, depending on target brightness. All previous observations of the targets in the same channels were retrieved from the \textit{Spitzer} Heritage Archive. Most targets have measurements at two or three epochs, and seven have between four and nine epochs in at least one channel. Identities and numbers of observations for each target are given in Table~\ref{tableResults}. Staring-mode observations of 1145+017 are excluded as they are not optimised for mosaicking, and have already been analysed \citep{Xu2018WD1145, Xu2019UVtransits1145}. Data in the 5.7 and {7.9\,\micron} channels from the cryogenic mission were not retrieved, as only five stars were observed more than once at these wavelengths, and the signal-to-noise (S/N) is not optimal for this sensitive study. 

Individual corrected basic calibrated data (CBCD) frames can be combined to create mosaics with 0.6\,arcsec\,pixel$^{-1}$, and both types of image are analysed here. Ready-made mosaics are provided in the archive, created by mission pipeline versions 18.12.0, 18.25.0, or 19.2.0, depending on the observation date. Mosaics are also built using the \textsc{mopex} package, following recommendations in the \textsc{mopex} User's Guide and the \textit{Spitzer} Data Analysis Cookbook.

Aperture photometry is performed on CBCD frames and \textsc{mopex} mosaics using the \textsc{apex} module, which also implements point-spread function (PSF) fitting\footnote{Strictly speaking, it performs point-response function fitting, but the concepts are sufficiently similar that the more familiar term is used here} for mosaics. Aperture photometry is performed on pipeline mosaics using the \textsc{apphot} module of \textsc{iraf}. Simulations using the IRAC~PSFs suggest that aperture radii of either two or three native pixels achieve optimal S/N, depending on target brightness. Therefore, in all cases, aperture radii of three native pixels are used, with sky annuli of 12--20~native pixels, achieving a good compromise between maximizing S/N and minimizing the impact of centroiding errors. The aperture corrections prescribed by the IRAC Instrument Handbook\footnote{Updated values for the warm mission are provided at \href{https://irsa.ipac.caltech.edu/data/SPITZER/docs/irac/calibrationfiles/ap_corr_warm/}{irsa.\hspace{0pt}ipac.\hspace{0pt}caltech.\hspace{0pt}edu/\hspace{0pt}data/\hspace{0pt}SPITZER/\hspace{0pt}docs/\hspace{0pt}irac/\hspace{0pt}calibrationfiles/\hspace{0pt}ap\_corr\_warm/}} are applied, but no colour corrections are made. The array-location-dependent corrections are applied, though for a target with good coverage near the centre of a mosaic they are usually negligible. Pixel-phase corrections are made only for CBCD frames, as the intrapixel variability effect averages out during mosaicking \citep{Wilson2019}. Uncertainties are calculated taking account of noise from target and sky, and the sky subtraction error, where the sky noise dominates at all but the brightest source. When absolute fluxes are used, calibration errors of 1.8\,per~cent~(3.6\micron) or 1.9\,per~cent~(4.5\micron) are added in quadrature \citep{Reach2005IRAC}, as is an aperture correction error of 2\,per~cent. However, the majority of the work in this study relies on relative fluxes.

In most cases the \textsc{apex} aperture photometry agrees with measurements made on the archive mosaics in \textsc{iraf}. Excluding a handful of outliers that differ by up to 15~per~cent, a Gaussian fit to the residuals has a width of 1.0~per~cent. The larger discrepancies are due either to neighbouring sources that cause a mismatch between centroids and PSF-fitted positions, or to inconsistent masking of cosmic rays. The most egregious example is at 0735+187, where two consecutive observations were made on MJD~55\,531. Cosmic rays fall within the target aperture in both channels during the first epoch, but are not masked in the archive mosaics, creating the illusion of a significant decrease in flux within two hours. Care was taken to deal with such issues in the \textsc{mopex} reductions.

\textit{Spitzer} IRAC photometry is stable to within 2~per~cent between measurements \citep {Reach2005IRAC}. Mosaics can achieve that level of precision, but CBCD images have lower S/N than mosaics by a factor of $\sqrt{n}$~frames. These limitations can be overcome by measuring the brightness of a source relative to field stars, i.e. differential photometry. Targets are typically located near the centre of mosaics, but sometimes they were observed incidentally in unrelated programs, and thus few comparison stars may be common to all images. For CBCD frames this problem is compounded by dithering, and sources are often not detected by \textsc{apex} in every frame. Inhomogeneous ensemble photometry is therefore employed, as it is designed to cope with these issues \citep{Honeycutt1992}. The technique determines the magnitude correction for each frame that minimizes the variance about the mean magnitude for all constant field stars. For the mosaics, residuals between raw \textsc{apex} measurements and differential measurements have a Gaussian distribution of width 1.7~per~cent, i.e. consistent with the 2~per~cent repeatability level.

\textsc{apex} performs both aperture photometry and PSF fitting for point sources. The former is the recommended procedure as the PSF is undersampled by IRAC, but the latter can be useful in crowded fields. Several targets have neighbouring sources that contaminate the photometric aperture, so PSF fitting is explored in those cases. However, the results are not sufficiently reliable for this study. The most crowded target is 1929$-$012, where two sources are known within about 2\,arcsec of the white dwarf from \textit{JHK} imaging, and several more lie within 5\,arcsec \citep{Melis2011}. \textsc{apex} does not identify all of those sources. Even when they are injected manually, not all survive the fitting procedure, and the number of sources retained varies between epochs. Since the work here focuses on relative fluxes rather than absolute fluxes, and the aim is to characterise the population rather than study individual objects in detail, aperture photometry remains acceptable even in crowded fields. In all cases, the majority of flux within the aperture derives from the target, and the risk of detecting variation from background objects is minimal: only around 1~per~cent of field sources are found to be variable in the mosaics studied here, consistent with previous IRAC studies \citep{Kozowski2010, Polimera2018}.
Thus, for all targets, the fluxes adopted for analysis are those obtained through differential photometry of \textsc{apex} aperture measurements.

Included in this study are observations of the binary system SDSS\,1557 made over two orbital periods of 2.3\,h, one in each channel (program~12106). A frame time of 100\,s was used in the medium cycling dither pattern, acquiring a total of 80 frames per channel. In addition to the procedures described above, aperture photometry is performed with \textsc{iraf} on the individual CBCD frames. Where cosmic rays impinge on the aperture the affected pixels are replaced with interpolated values, where possible, otherwise the frame is discarded.

Four stars are known to be ZZ\,Ceti pulsators: 1116+026, 1150$-$153, 1541+651, and 2326+049. Lomb--Scargle periodograms are computed for other stars in the sample from optical photometric survey data, where available (CRTS, \citealt{Drake2009}; PTF, \citealt{Law2009}; TESS, \citealt{Ricker2014}; ZTF, \citealt{Masci2018}). Significant peaks are found near a period of 3270\,s for 1145+288, and reported here for the first time. The stellar parameters lie within the ZZ\,Ceti instability strip ($T_{\text{eff}}=12$\,$140\pm210$\,K, $\log{g}=8.14\pm0.10$; \citealt{Xu2019composition}), but the available data do not permit a firm identification, as the period found is longer than expected for that stellar class \citep{Hermes2017}. Nevertheless, it is labelled as a pulsator for the purposes of this study, as it appears to be intrinsically variable.

\section{Results}
\label{sectionResults}

\begin{figure*}
\includegraphics[width=0.975\columnwidth]{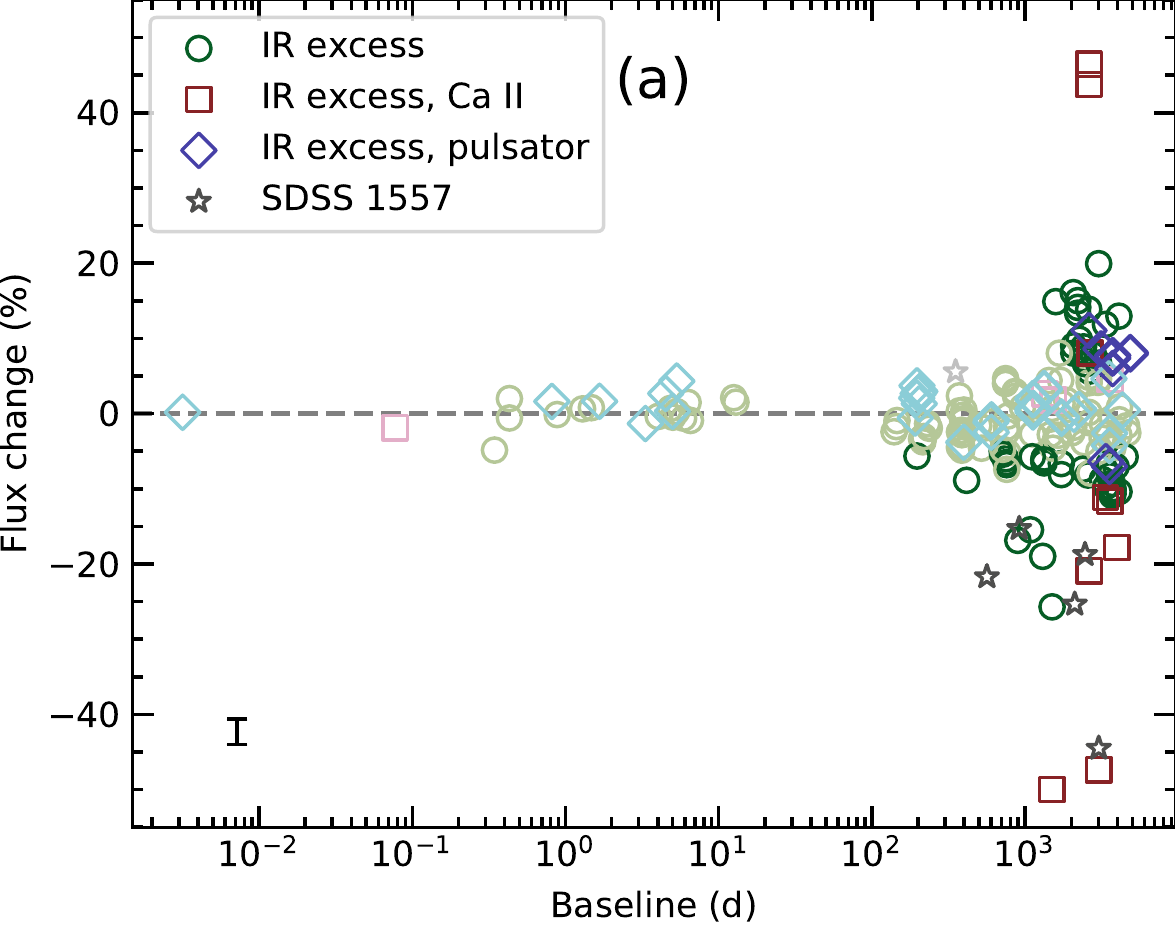}
\hspace{0.05\columnwidth}
\includegraphics[width=0.975\columnwidth]{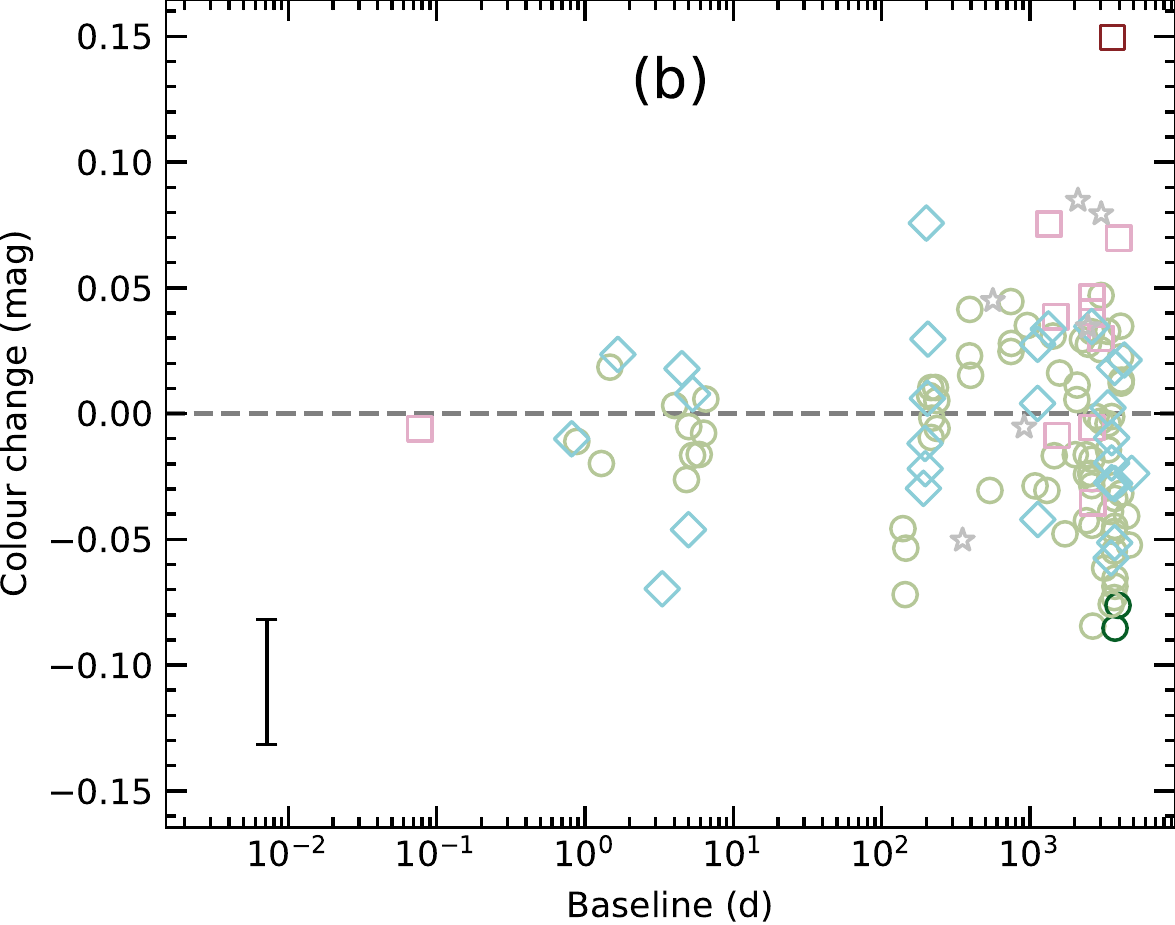}\\
\vspace{0.05\columnwidth}
\includegraphics[width=0.975\columnwidth]{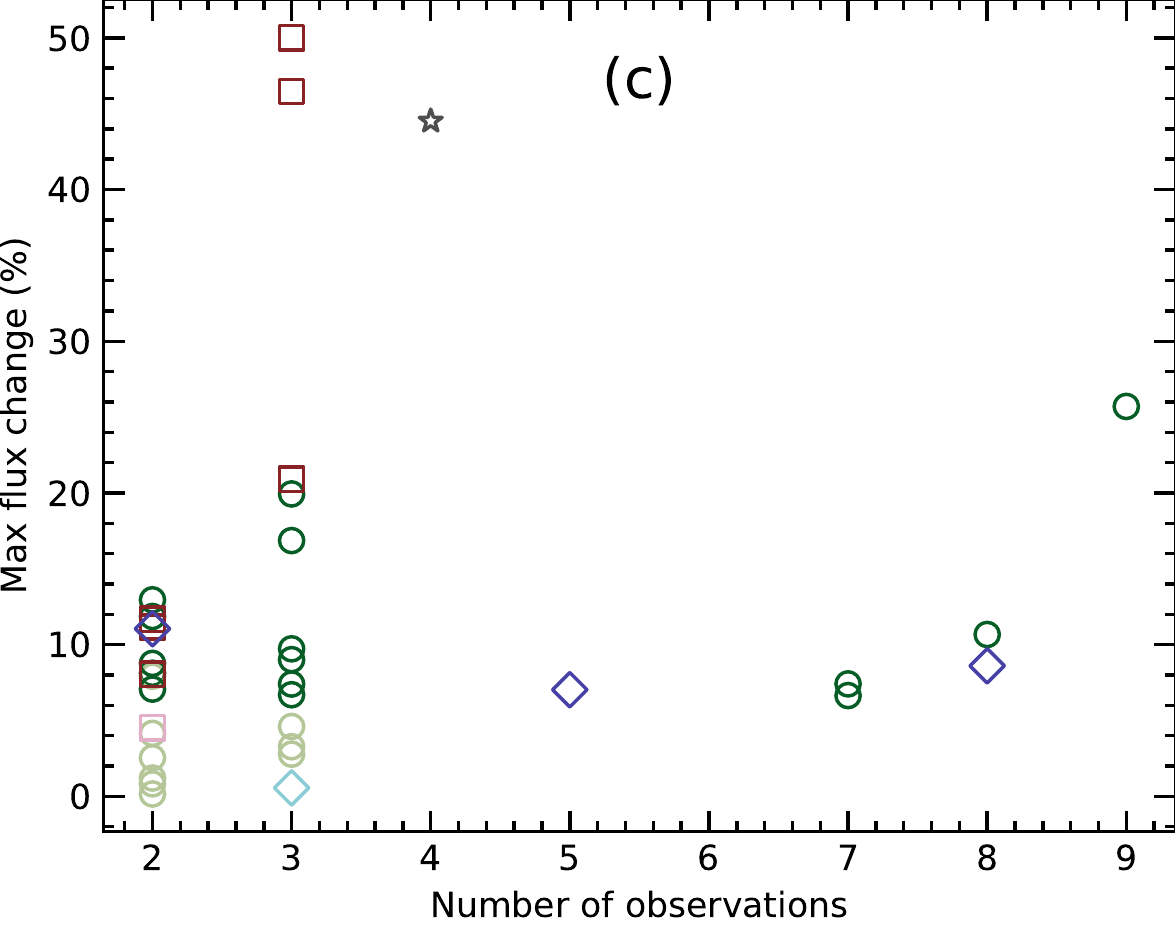}
\hspace{0.05\columnwidth}
\includegraphics[width=0.975\columnwidth]{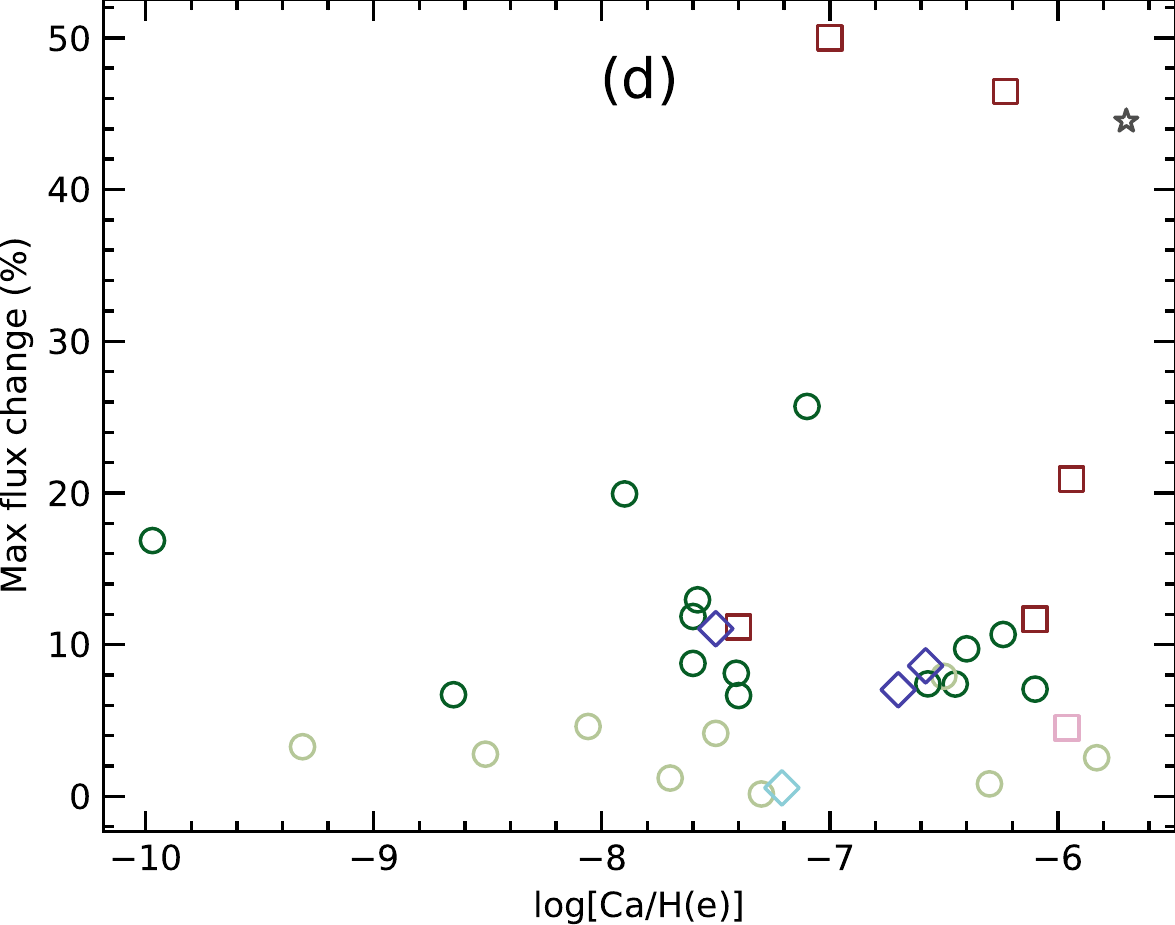}
\caption{(a) Flux changes versus time baseline for all pairs of {4.5\,\micron}~measurements. (b) Change in colour versus time baseline for all pairs of measurements where both channels are available. (c) Maximum relative {4.5\,\micron} flux changes for each star versus number of observations. (d) Maximum relative {4.5\,\micron} flux changes versus Ca abundance. Darker symbols indicate significance above $3\upsigma$. Median uncertainty shown at lower left, where appropriate.}
\label{figureMain}
\end{figure*}

Fractional flux changes quoted throughout are calculated with respect to the lowest flux value, irrespective of the direction of the change, i.e. $\Delta f=(f_2-f_1)/\text{min}(f_1,f_2)$. Variation can be seen across the sample. The median and maximum peak-to-peak variations at {3.6\,\micron} are 5~per~cent and 56~per~cent, and at {4.5\,\micron} they are 8~per~cent and 50~per~cent. The 3$\upsigma$ detection limit for variation has a median around 5~per~cent and rises to 19~per~cent for the faintest targets. Figure~\ref{figureMain}(a) plots {4.5\,\micron}~flux changes against the time between measurements for all pairs of epochs at each star, and encapsulates most of the salient features of this study. Peak-to-peak variations are also given in Table~\ref{tableResults}.

Figure~\ref{figureMain}(b) shows colour changes against the time between measurements for all available pairs of measurements, where colour is the ratio between {3.6\,\micron} and {4.5\,\micron} fluxes. Colour remains approximately constant, and in most cases any difference is below 3$\upsigma$ significance. For this reason only results from {4.5\,\micron} data are plotted throughout this study, as results are similar in both channels.

Three interesting features emerge from panels (a) and (b) of Figure~\ref{figureMain}. First, the largest changes tend to be seen at systems where \ion{Ca}{ii} triplet emission has been reported. Second, pulsators appear to behave similarly to the wider sample. Third, there is no obvious preferred direction for flux changes: brightening is as common as dimming.

The gap in the $\upDelta t$~distribution results from operational limits on the sun-telescope angle, such that most targets only become accessible to \textit{Spitzer} during two observing windows of around 40\,d each year. No significant variation is detected on time-scales of hours to weeks, with only one exception. A low-amplitude change at {3.6\,\micron} is seen at 2326+049, a bright pulsator, and pulsation time-scales are typically below one hour \citep{Hermes2017}. Significant changes at non-pulsating stars appear immediately beyond the gap, starting from $\upDelta t=144$~d (an increase in the {3.6\,\micron} flux at 0408$-$041). Time-scales for stars with \ion{Ca}{ii}~emission are poorly sampled, but their flux changes appear to have a similar distribution to the wider population, yet with higher variance. Larger changes tend to occur over longer baselines. Were such changes due to frequent but transient events then the odds of detecting high variation would increase with the number of observations. Figure~\ref{figureMain}(c) plots maximum flux change against number of epochs for each star, verifying that the correlation between larger changes and longer baselines is real.

Figure~\ref{figureMain}(d) shows the largest variation for each target against the Ca abundance reported in the literature, where available. There is a weak (p-value~0.1) correlation between flux changes and abundances. However, this appears only to trace population membership: when \ion{Ca}{ii}~emission stars are separated from the rest of the sample, there is no significant correlation within either subset. Stellar parameters, atmosphere type, and distance are uncorrelated with infrared variation, as expected.

\begin{figure}
 \includegraphics[width=\columnwidth]{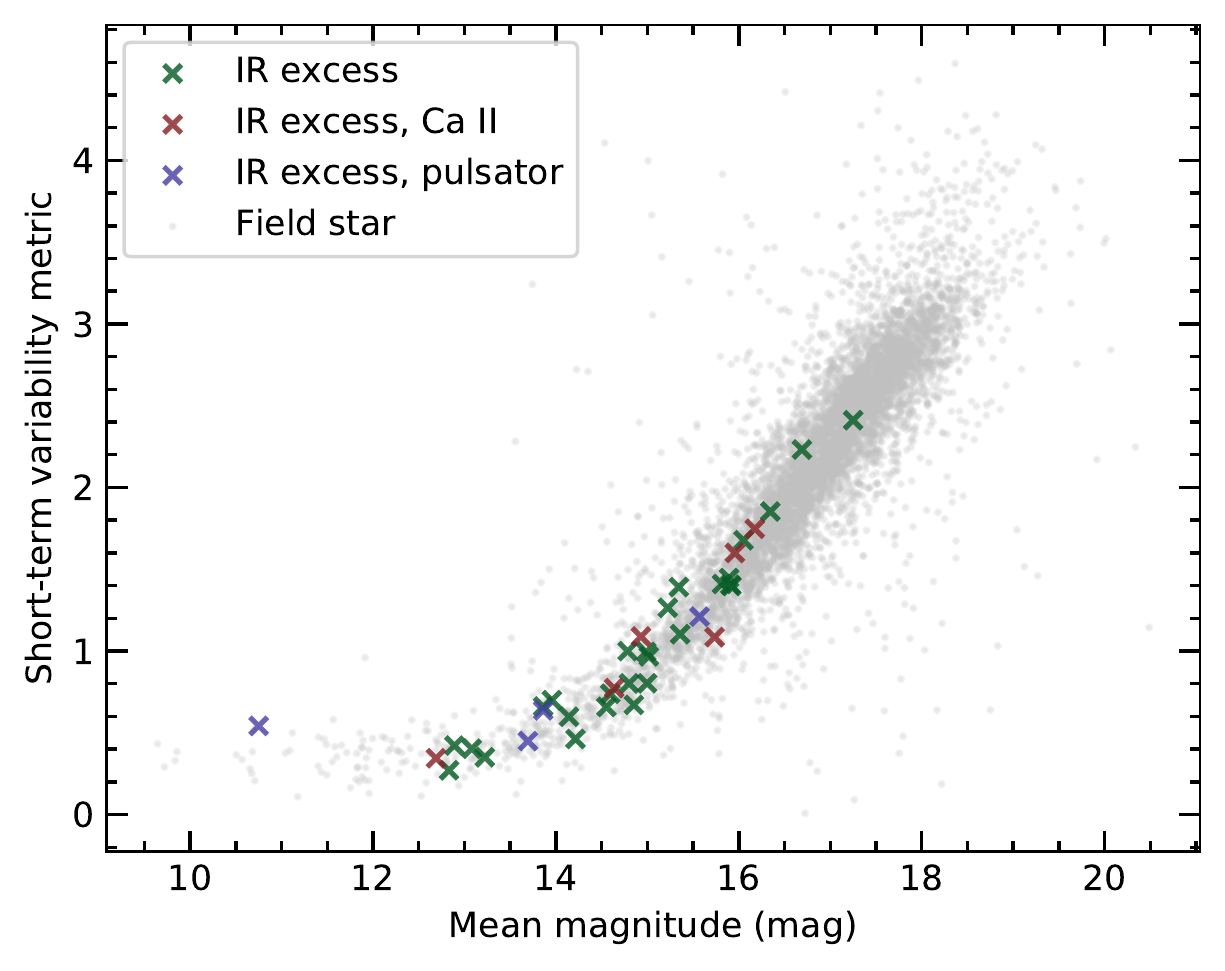}
 \caption{Short-term variability, according to the metric defined in Eqn.~\ref{equationScatter}, plotted against mean magnitude. For any given source, the variability metric quantifies the scatter in epoch-mean-subtracted {4.5\,\micron} flux across all individual exposures from which mosaics are assembled. Variable sources manifest as points lying above the main locus of stable sources. Crosses show dusty white dwarfs, with \ion{Ca}{ii}~emission stars in red and pulsators in blue, while field sources are grey points.}
 \label{figureShortBaseline}
\end{figure}

To search for variability on the shortest time-scales, the scatter in flux measurements in the CBCD frames is examined. As most sources in the field are expected to be photometrically stable, their measurement scatter should increase with faintness, and variable sources will manifest as outliers above the trendline. The quantity of interest is the scatter in residual fluxes after subtracting epoch means. However, as exposure times vary between observations, the residuals are scaled so that each frame has the same effective photon noise level. Specifically, the short-term scatter $S$ in flux~$f$ for a given star is calculated by adapting the standard deviation formula as follows:

\begin{equation}
S=\sqrt{\frac{\sum_{j=1}^{N}\sum_{i=1}^{n_j}(\phi_{j,i}(f_{j,i}-\bar{f_{j}}))^{2}}{(\sum_{j=1}^{N}n_j)-1}}
\label{equationScatter}
\end{equation}

where there are $N$~observations of $n$~frames each, and where gain~$G$, exposure time~$t$, and flux conversion factor~$C$ are used to find the scaling factor $\phi=\sqrt{Gt/C}$. Results for the {4.5\,\micron} channel are shown in Figure~\ref{figureShortBaseline}, where the dusty white dwarfs lie on the same locus as field stars, and a similar result is seen at {3.6\,\micron}. As a further test, the target light curves derived from the CBCD frames in each observation are fitted with straight lines. The resulting gradients show normally-distributed deviations from their mean of zero, with only two outliers, at SDSS\,1557 (as expected; see Section~\ref{subsectionSDSS1557results}) and at 2326+049 (a pulsator). Finally, Lomb--Scargle periodograms are constructed from the \textit{Spitzer} data for each star, and no significant peaks are found. There is thus no evidence for variation in dust emission on time-scales of minutes to hours.

\subsection{Individual objects}
\label{subsectionIndividualObjects}

Three stars in the sample have not previously shown an excess at {3.6\,\micron}, namely 1225$-$079, 1455+298, and 2132+096. The {3.6\,\micron} flux at the latter star has increased by 6\,per\,cent, confirming the change noted from \textit{WISE} data, albeit with a smaller magnitude \citep{Swan2019}. However, the binned \textit{WISE} data have $\text{S/N}\leq6$, and moreover the recent \textit{Spitzer} measurement was made 56\,d prior to the closest \textit{WISE} epoch, so a direct comparison cannot be made. The other two stars remain stable within their errors in the data examined here, and are sufficiently bright to allow ongoing monitoring by \textit{WISE}; while 1455+298 had hitherto been contaminated by a background source, its proper motion of 615\,mas\,yr\textsuperscript{-1} has isolated it sufficiently that measurements in future data releases should be reliable.

\begin{figure}
 \includegraphics[width=\columnwidth]{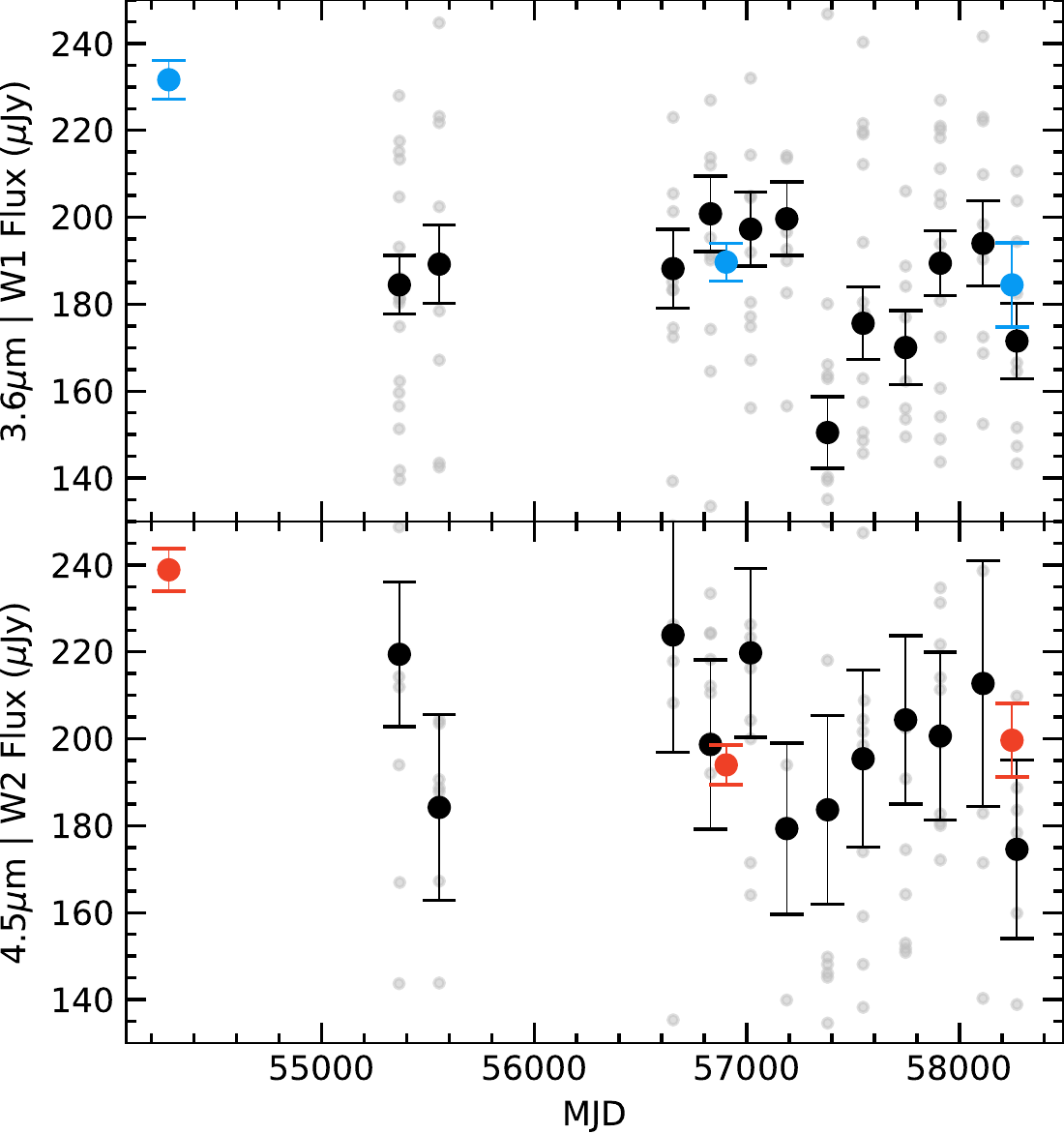}
 \caption{Flux measurements at 1226+110. Coloured points are from \textit{Spitzer}~IRAC, while \textit{WISE} data are shown in grey (individual measurements) and black (weighted means).}
 \label{figure1226}
\end{figure}

IRAC measurements are unchanged between the last two epochs at 1226+110, a \ion{Ca}{ii} emission system, but the apparent stability is deceptive. Data from \textit{WISE} are retrieved from the 2019 data release and weighted averages of each epoch are taken, following \cite{Swan2019}. A sharp decrease in {3.6\,\micron} flux can be seen around MJD~57\,300 in Figure~\ref{figure1226}. It is thus clear that consistency between \textit{Spitzer} measurements separated by years cannot be taken as evidence for stability, a point underlined by the changes on time-scales of months found here for some stars.

\begin{figure}
 \includegraphics[width=\columnwidth]{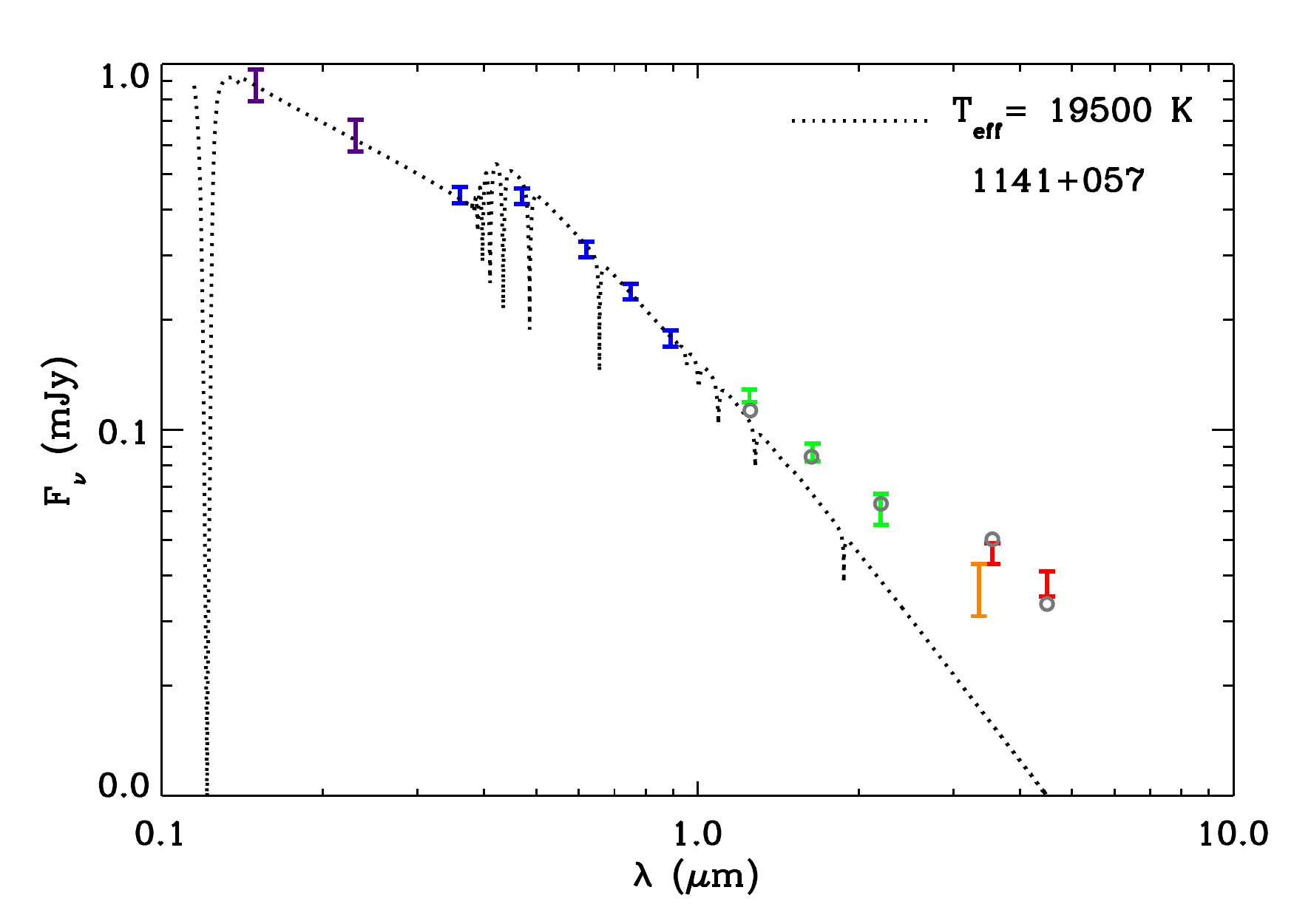}
 \caption{SED of 1141+057 showing photometry from \textit{GALEX} (purple), SDSS (blue), 2MASS (green), \textit{WISE} (orange), and \textit{Spitzer} (red). The dotted line shows the stellar model \citep{Koester2010}, while circles represent the addition of a type~L5 dwarf \citep{Patten2006} with estimated day-side illumination.}
 \label{figure1232}
\end{figure}

\textit{Spitzer} data had not previously been taken or published for eight stars, namely 0420$-$731, 0536$-$479, 0842+572, 1141+057, 1145+288, 1232+563, 1536+520, and 2329+407. All are now confirmed to show real infrared excesses, based on examination of mosaic images and spectral energy distributions~(SED). Background sources that lie within the \textit{WISE} PSF are visible in the mosaics for most of these targets. However, contamination of \textit{WISE} flux measurements significantly above the per\,cent level is only likely at 0536$-$479, as was already known for that star \citep{Dennihy2016}. The excesses are consistent with emission from warm dust at all of these stars but one: 1141+057 appears to have a low-mass companion. A reasonable fit to the SED can be obtained if the star is accompanied by an ultra-cool dwarf of spectral type~L3, or of type~L5 that is irradiated by the white dwarf. Figure~\ref{figure1232} shows observed fluxes together with those predicted under the latter model, where the contribution from day-side illumination of the companion is estimated by reference to a similar binary, 0137$-$349 \citep{Casewell2015}. It must be emphasized that such a model is heuristic, as the properties of the companion and the orbital phases of \textit{Spitzer} observations are unknown. The star had been identified previously as potentially hosting a debris disc on the basis of its infrared flux and the presence of \ion{Ca}{ii} emission \citep{Guo2015}. The emission lines are now known to show radial velocity variation consistent with origin in a low-mass companion \citep{Florez2020}, and thus there is no evidence for a warm dust disc in the system, which appears instead to be a binary.

\subsection{\texorpdfstring{SDSS\,J155720.77+091624.6}{SDSS J155720.77+091624.6}}
\label{subsectionSDSS1557results}

\begin{figure}
 \includegraphics[width=\columnwidth]{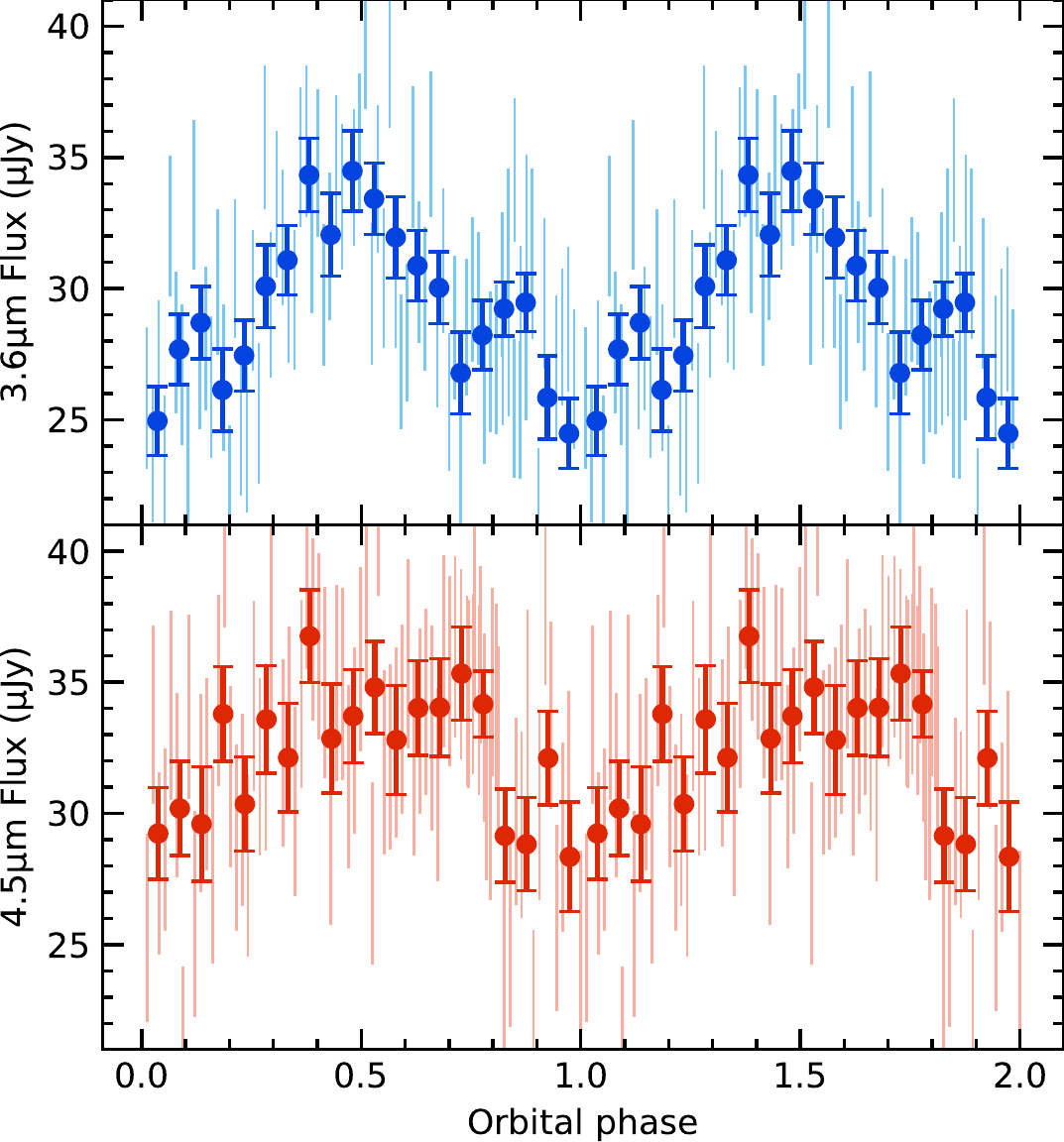}
 \caption{IRAC light curve for SDSS\,1557 (program~12106 only), phase-folded to the binary ephemeris and duplicated over two orbital periods, with {3.6\micron} data in blue and {4.5\micron} in red. Individual measurements are shown in the background as vertical bars whose extents are the uncertainties, and are binned to produce the points with error bars.}
 \label{figureSDSS1557}
\end{figure}

This unusual system, where warm dust surrounds a white dwarf and its substellar companion, has been observed by \textit{Spitzer} on four occasions over 8\,yr. One observation acquired sufficient frames to assemble a light curve over a full orbital period in each channel, as shown in Figure~\ref{figureSDSS1557}. Although the data are noisy (typical $\text{S/N}=8$), an approximately sinusoidal component is apparent and consistent with the spectroscopically determined binary period of 2.3\,h \citep{Farihi2017SDSS1557}. When phase-folded against the binary ephemeris, the periodic infrared variation is in antiphase with the radial velocity, almost certainly due to dayside--nightside effects of stellar irradiation.

There are also flux changes on longer time-scales and with larger amplitude than the periodic variation, and thus they must have a distinct origin, i.e. in the circumbinary dust. Long-term variation in the dust emission can be disentangled from the periodic signal by jointly fitting all light curve segments and radial velocity measurements, treating them as sinusoids in antiphase with a common period. After removing the periodic signal, significant variation in the dust emission on yearly time-scales remains, and indeed the peak-to-peak fractional change increases slightly to become larger than any other in this study.

A toy model of the irradiated companion where both objects are treated as blackbodies, and tuned so that the {3.6\,\micron} light curve matches the observations, predicts higher periodic variation at {4.5\,\micron} than at shorter wavelengths. Such behaviour is observed at another white dwarf--L~dwarf binary (0137$-$349; \citealt{Casewell2015}). However, the opposite is true at SDSS\,1557, where the semi-amplitudes of the 3.6 and {4.5\,\micron} signals are 11 and 8~per~cent, respectively (compared to approximately 11 and 15~per~cent at 0137$-$349). The discrepancy between model and data is likely due to dilution of the companion flux by dust emission, but molecular absorption and clouds may also cause the companion SED to depart substantially from a blackbody. However, without \textit{JHK} light curves to complement the \textit{Spitzer} data and constrain the system parameters, these two possibilities cannot confidently be distinguished.

\section{Discussion}
\label{sectionDiscussion}

As a population, dusty white dwarfs are known from \textit{WISE} data to vary in the infrared \citep{Swan2019}. Using the higher spatial resolution and sensitivity of \textit{Spitzer}, this study confirms that result at the level of individual stars, and indicates that collisional processes are responsible. Several key findings, and the data and rationale that support them, are discussed in this section. For convenience, they are summarised as follows:

\begin{enumerate}
  \item Infrared variation appears ubiquitous at dusty white dwarfs, where brightening and dimming events are equally common, and significant colour changes are minimal.\\
  
  \item Variation manifests over time-scales of months to years, with a trend towards larger changes over longer baselines. No variation is seen on time-scales of minutes to days, save at pulsating stars.\\

  \item Canonical, flat, and vertically optically thick discs are insufficient to account for the observed variation, although their presence cannot be ruled out.\\

  \item The greatest flux changes are seen at \ion{Ca}{II} emission systems, a strong indication that ongoing collisions are responsible for the observed gas and dust.
\end{enumerate}

\subsection{Characteristics of variation}
\label{subsectionCharacterisingVariation}

Significant variation is detected at nearly three-quarters of the stars, and is thus likely to be ubiquitous. This is despite 80~per~cent of the sample being observed at only two or three epochs with effectively random cadences. The underlying mechanisms appear to be universal, as flux changes are not correlated with stellar parameters, or with photospheric Ca abundances\footnote{Strictly, there is a weak correlation, but it is incidental to the point at hand, as it reflects the higher Ca abundances and variability of \ion{Ca}{ii} emission stars as a population (see Section~\ref{sectionResults}). A direct link between Ca abundance and infrared variability is invalidated by the stars with high Ca abundance where \ion{Ca}{ii} emission is known to be absent \citep{Manser2020}.} and thus average or instantaneous accretion rate. The latter point is potentially an important clue as to dust configuration, as accretion rates should correlate with infrared emission for optically thin dust, but not where it is optically thick \citep{Bonsor2017}.

The stability in colour rules out large changes in temperature profile of the emitting material, such as might result from a dramatic change in radial extent of a disc, motion along an eccentric orbit, or orbital evolution through PR drag. Geometric changes are not entirely ruled out as a cause of variation, and there is strong evidence that the configuration of circumstellar debris can change. For example, observations of several systems are consistent with precession of an eccentric ring of material \citep{Manser2016a, Dennihy2018, Cauley2018}, in line with theoretical expectations \citep{Miranda2018}. However, it is not clear that a flat disc with fixed inclination to the observer can result in flux variation, unless occulted by the star. A warped disc might change in brightness if the disc geometry is time-dependent, possibly in a periodic manner, but there is no evidence for this in the sparse data analysed here.

There is a lack of a net direction for flux changes across the entire sample. Under binomial statistics, the signs of flux changes show no significant departure from a random distribution with equal probabilities, and Figure~\ref{figureMain}A shows that this holds across all time-scales. The cumulative magnitude of fractional variation is negative in both channels, though a few large dimming events at \ion{Ca}{ii} systems dominate this result, and a bootstrapping analysis finds it to be significant only at the $1\upsigma$ level. In a dataset that is temporally well sampled, these data would disfavour dust production and removal operating on different time-scales. However, with the current limited number and frequency of \textit{Spitzer} epochs, such a scenario cannot be ruled out.

A good example of poorly-characterised, bidirectional flux changes is the iconic dimming event at 0956$-$017 \citep{Xu2014drop}. While it was suggested that the system lost a significant radial component of a canonical disc, the mid-infrared colour stayed constant within the uncertainties. NEOWISE data indicate that the flux later recovered \citep{Swan2019}, implying an alternative process. By the time of the latest \textit{Spitzer} measurement the flux had once again returned to a low state. However, the low S/N data from NEOWISE do not constrain the epoch or duration of the brightening episode.

\subsection{Time-scales of changes}
\label{subsectiontime-scales}

The only significant flux changes on time-scales shorter than months are seen at pulsating stars. However, the apparent absence of variation in dust emission over minutes to hours may simply reflect sparse temporal coverage. Almost all the stars have accrued less than an hour of total exposure time across a few tens of frames, which are used here to probe the shortest time-scales. Likewise, only eight have been observed at epochs spaced less than a week apart. Thus, while the data offer no support for changes on orbital time-scales within a canonical disc (e.g. rapid re-incorporation of impact-generated debris clouds; \citealt{Farihi2018GD56}), transient dimming or brightening events with low duty cycles cannot be ruled out.

The first measurements that show a significant variation are 144\,d apart, almost the shortest baseline beyond the typical \textit{Spitzer} target visibility window. The trend to larger variation on longer, year to decade time-scales is potentially an indicator of orbiting planet(esimal)s or their disrupted fragments. If dust-producing bodies are distributed inhomogeneously along their orbit, as in recent simulations of tidal disruption events (\citealt{Malamud2020A}; see also \citealt{Hahn1998}), the time-scale of infrared variation may be an indicator of the orbital period. For example, as discussed by \cite{Nixon2020}, an eccentric ring of debris may have a semi-major axis that implies the bulk of dust is too cool for detection, yet passes sufficiently close to the star to cause 1000\,K dust emission. If that dust is distributed non-uniformly along its orbit, there might be both geometrical and temporal changes in the emission detected by \textit{Spitzer} and \textit{WISE}. For the foreseeable future, it may be the case that only \textit{WISE} can constrain the relevant time-scales of orbits, as it is the sole space-based facility in operation at the wavelengths of interest.

\subsection{The canonical disc re-evaluated}
\label{subsectionCanonicalDisc}

To account for dusty white dwarf properties, a circular, geometrically thin, and optically thick annulus of dust is often invoked \citep{Farihi2016}. It is argued here that this canonical disc -- first suggested by \cite{Jura2003} -- is neither necessary nor sufficient, and that the presence of optically thin dust is required. On the one hand, the canonical disc has theoretically appealing properties, yet cannot easily show infrared variation. On the other hand, optically thin dust presents modelling challenges, but is consistent with observations. These issues are discussed below, and it is important to note that the two configurations can coexist.

All infrared observations to date only strictly require the presence of optically thin dust: emission from micron-sized particles can produce the observed infrared excesses, and changes in the emitting surface area can produce the observed variation. Moreover, it is well established that some systems \textit{must} host such dust, for example to account for the silicate emission features observed near {10\,\micron} \citep{Jura2007GD362, Farihi2017SDSS1557, Farihi2018GD56}. Some theoretical models predict that all dust may be optically thin \citep{Bonsor2017}, although they cannot yet account for the frequency and brightness of infrared excesses. Such discs require replenishment on PR~drag time-scales, which are typically less than 10\,yr for micron-sized dust \citep{Farihi2016}. Notably, 1000\,K dust emission persists for the entire sample over baselines exceeding a decade. At some systems, unshielded blackbody grains at this temperature would lie outside the Roche limit \citep{Jura2007Spitzer11stars, Farihi2018GD56}. Despite these challenges, there are some advantages to models with optically thin dust. Simulations starting from a narrow and mildly eccentric annulus of dust particles produce a collisional cascade that maintains a large vertical scale height \citep{Kenyon2017collisions}. The resultant infrared excesses in these models fluctuate between high and low states, where variation can occur over a wide range of time-scales, potentially consistent with the data presented here. Thus, production and destruction of optically thin dust offers a promising framework for the observations.

In contrast, the utility of the canonical disc model includes -- but is not limited to -- a lifetime that greatly exceeds that of PR~drag, and consistency with dynamical relaxation on orbital time-scales of hours within the stellar Roche limit \citep{Jura2003}. Importantly, observed infrared excesses can typically be modelled by discs located between the Roche limit and the region in which solids rapidly sublimate, in line with expectations if the emitting debris results from tidal disruptions \citep{Rocchetto2015}. Furthermore, an opaque disc can harbour sufficient mass to deliver the metal masses exceeding $10^{23}$\,g that are observed in some white dwarf photospheres (e.g.~\citealt{Dufour2010}), orders of magnitude above the limit of an equivalent, optically thin disc ($\sim10^{19}$\,g).

However, it has become clear that the canonical disc is not the complete picture. It is a poor fit for the disc at 0408$-$041 \citep{Jura2007GD362, Farihi2018GD56}, and the brightness of discs inferred at systems with \ion{Ca}{ii} emission is at odds with their inclinations \citep{Dennihy2017}. The circumbinary disc orbiting SDSS\,1557 is a clear case where a canonical configuration is impossible, as it is dynamically prohibited \citep{Farihi2017SDSS1557}. In particular, based on this study, the observed infrared variations are challenging or impossible to account for by an unwarped canonical disc alone, as there are as yet no changes on the expected orbital time-scales. At this stage, the $T\sim1000$\,K dust that persists at white dwarfs could be in a canonical configuration or otherwise, but on the basis of all \textit{Spitzer} and \textit{WISE} data analysed to date, it is hypothesised that white dwarf debris discs possess substantial components of optically thin dust, possibly in addition to canonical dust, where collisions play a dominant role in the overall evolution.

\subsection{The link between detected gas and infrared fluctuation}
\label{subsectionGas}

The higher variability of \ion{Ca}{ii}~emission systems firmly establishes a link between the production of gas and dust. The gas emission systems are indistinct from the other targets in terms of \textit{Spitzer} observations (number, cadence, exposure length, S/N). Thus, they only stand apart via their higher infrared variability. The most straightforward and compelling interpretation is that they represent the high tail of a distribution of gas production rates that typically remain below the threshold of excitation, stability, and thus detection. In this picture, they are more dynamically active than their peers, experiencing more prolific collisions, possibly via fragments resulting from a recent tidal disruption event. While forming an observationally distinct class, \ion{Ca}{ii}~emission systems may possess debris discs similar to those where only dust is observed, the only difference being their present level of activity.

Although not directly related to the infrared data presented here, a plausible scenario for collisional gas production in \ion{Ca}{ii} emission systems is considered qualitatively below. Collisions in a canonical disc should be efficiently damped and unimportant \citep{Metzger2012}. However, for a body that is not exactly co-orbital with circularised dust near the stellar Roche limit, even a small eccentricity of $e\sim0.01$ can result in shear velocities of several~\kms. Impacts by dust grains on the surface of such a body are likely to produce gas at these velocities \citep{Tielens1994}. While dust discs reside within the Roche limit, fragments below around 1\,km in size with internal strength can resist disruption, and thus may participate in gas production \citep{Brown2017}. Any orbit that intersects the disc can produce collisions, with the potential for gas production. The planetesimal orbiting within the debris disc at 1226+110 is one example of such a configuration \citep{Manser2019}.

\subsection{Outlook}
\label{subsectionOutlook}

\textit{Spitzer} and \textit{WISE} have revealed the ubiquitous infrared variability of dusty white dwarfs, where collisional processes are implicated, but substantial observational gaps remain to be filled. Continued \textit{K}-band monitoring will be important to reach the decadal time-scales over which the largest variation has been observed, and to probe the hottest circumstellar dust. At wavelengths beyond {5\,\micron}, coverage is sparse, and variability thus unexplored here, but such data is required to probe more distant dust and to constrain disc models. Short-cadence monitoring may reveal or rule out transient phenomena on the hourly time-scales of orbital periods near the Roche limit. In terms of targets, polluted stars that have shown no infrared excess in previous surveys should not be overlooked, as this study finds that significant dust production can occur within a year.

Theoretical advances will be required to interpret these data. For example, it is unclear how the highly eccentric debris rings thought to form in tidal disruptions evolve towards the closely-orbiting discs that are observed, and indeed whether a canonical disc ever results \citep{Nixon2020}. The enhanced activity at \ion{Ca}{ii} emission systems makes them promising targets for testing models, especially those invoking collisions.

\section*{Acknowledgements}

The authors are grateful to referee Sandy Leggett for making several suggestions that improved the paper, and to Chris Nixon and Jim Pringle for extensive discussions. This work is based on observations made with the \textit{Spitzer Space Telescope}, which was operated by the Jet Propulsion Laboratory (JPL), California Institute of Technology (CIT) under a contract with NASA. This publication makes use of data products from the \textit{Wide-field Infrared Survey Explorer}, a joint project of the University of California, Los Angeles and JPL/CIT, and NEOWISE, a project of JPL/CIT, both funded by NASA. AS~acknowledges support from a Science and Technology Facilities Council (STFC) studentship. JF~acknowledges support from STFC grant ST/R000476/1. SGP~acknowledges the support of an STFC Ernest Rutherford Fellowship. TGW~acknowledges support from STFC grant ST/R000824/1.

\section*{Data availability statement}
The data used in this article are available in the Spitzer Heritage Archive at \href{https://sha.ipac.caltech.edu/applications/Spitzer/SHA/}{sha.\hspace{0pt}ipac.\hspace{0pt}caltech.\hspace{0pt}edu/\hspace{0pt}applications/\hspace{0pt}Spitzer/\hspace{0pt}SHA/}




\bibliographystyle{mnras}
\bibliography{Infrared_variability_II} 

\begin{thebibliography}{}
\makeatletter
\relax
\def\mn@urlcharsother{\let\do\@makeother \do\$\do\&\do\#\do\^\do\_\do\%\do\~}
\def\mn@doi{\begingroup\mn@urlcharsother \@ifnextchar [ {\mn@doi@}
  {\mn@doi@[]}}
\def\mn@doi@[#1]#2{\def\@tempa{#1}\ifx\@tempa\@empty \href
  {http://dx.doi.org/#2} {doi:#2}\else \href {http://dx.doi.org/#2} {#1}\fi
  \endgroup}
\def\mn@eprint#1#2{\mn@eprint@#1:#2::\@nil}
\def\mn@eprint@arXiv#1{\href {http://arxiv.org/abs/#1} {{\tt arXiv:#1}}}
\def\mn@eprint@dblp#1{\href {http://dblp.uni-trier.de/rec/bibtex/#1.xml}
  {dblp:#1}}
\def\mn@eprint@#1:#2:#3:#4\@nil{\def\@tempa {#1}\def\@tempb {#2}\def\@tempc
  {#3}\ifx \@tempc \@empty \let \@tempc \@tempb \let \@tempb \@tempa \fi \ifx
  \@tempb \@empty \def\@tempb {arXiv}\fi \@ifundefined
  {mn@eprint@\@tempb}{\@tempb:\@tempc}{\expandafter \expandafter \csname
  mn@eprint@\@tempb\endcsname \expandafter{\@tempc}}}

\bibitem[\protect\citeauthoryear{Barber, Patterson, Kilic, Leggett, Dufour,
  Bloom  \& Starr}{Barber et~al.}{2012}]{Barber2012}
Barber S.~D.,  Patterson A.~J.,  Kilic M.,  Leggett S.~K.,  Dufour P.,  Bloom
  J.~S.,   Starr D.~L.,  2012, \mn@doi [ApJ] {10.1088/0004-637X/760/1/26}, 760,
  26

\bibitem[\protect\citeauthoryear{Barber, Kilic, Brown  \& Gianninas}{Barber
  et~al.}{2014}]{Barber2014}
Barber S.~D.,  Kilic M.,  Brown W.~R.,   Gianninas A.,  2014, \mn@doi [ApJ]
  {10.1088/0004-637X/786/2/77}, 786, 77

\bibitem[\protect\citeauthoryear{Becklin, Farihi, Jura, Song, Weinberger  \&
  Zuckerman}{Becklin et~al.}{2005}]{Becklin2005}
Becklin E.~E.,  Farihi J.,  Jura M.,  Song I.,  Weinberger A.~J.,   Zuckerman
  B.,  2005, \mn@doi [ApJ] {10.1086/497826}, 632, L119

\bibitem[\protect\citeauthoryear{Bergfors, Farihi, Dufour  \&
  Rocchetto}{Bergfors et~al.}{2014}]{Bergfors2014}
Bergfors C.,  Farihi J.,  Dufour P.,   Rocchetto M.,  2014, \mn@doi [MNRAS]
  {10.1093/mnras/stu1565}, 444, 2147

\bibitem[\protect\citeauthoryear{Bonsor, Farihi, Wyatt  \& {Van
  Lieshout}}{Bonsor et~al.}{2017}]{Bonsor2017}
Bonsor A.,  Farihi J.,  Wyatt M.~C.,   {Van Lieshout} R.,  2017, \mn@doi
  [MNRAS] {10.1093/mnras/stx425}, 468, 154

\bibitem[\protect\citeauthoryear{Brinkworth, G{\"{a}}nsicke, Marsh, Hoard  \&
  Tappert}{Brinkworth et~al.}{2009}]{Brinkworth2009}
Brinkworth C.~S.,  G{\"{a}}nsicke B.~T.,  Marsh T.~R.,  Hoard D.~W.,   Tappert
  C.,  2009, \mn@doi [ApJ] {10.1088/0004-637X/696/2/1402}, 696, 1402

\bibitem[\protect\citeauthoryear{Brinkworth, G{\"{a}}nsicke, Girven, Hoard,
  Marsh, Parsons  \& Koester}{Brinkworth et~al.}{2012}]{Brinkworth2012}
Brinkworth C.~S.,  G{\"{a}}nsicke B.~T.,  Girven J.~M.,  Hoard D.~W.,  Marsh
  T.~R.,  Parsons S.~G.,   Koester D.,  2012, \mn@doi [ApJ]
  {10.1088/0004-637X/750/1/86}, 750, 86

\bibitem[\protect\citeauthoryear{Brown, Veras  \& G{\"{a}}nsicke}{Brown
  et~al.}{2017}]{Brown2017}
Brown J.~C.,  Veras D.,   G{\"{a}}nsicke B.~T.,  2017, \mn@doi [MNRAS]
  {10.1093/mnras/stx428}, 468, 1575

\bibitem[\protect\citeauthoryear{Casewell et~al.,}{Casewell
  et~al.}{2015}]{Casewell2015}
Casewell S.~L.,  et~al., 2015, \mn@doi [MNRAS] {10.1093/mnras/stu2721}, 447,
  3218

\bibitem[\protect\citeauthoryear{Cauley, Farihi, Redfield, Bachman, Parsons  \&
  G{\"{a}}nsicke}{Cauley et~al.}{2018}]{Cauley2018}
Cauley P.~W.,  Farihi J.,  Redfield S.,  Bachman S.,  Parsons S.~G.,
  G{\"{a}}nsicke B.~T.,  2018, \mn@doi [ApJ] {10.3847/2041-8213/aaa3d9}, 852,
  L22

\bibitem[\protect\citeauthoryear{Coughlin \& Nixon}{Coughlin \&
  Nixon}{2015}]{Coughlin2015}
Coughlin E.~R.,  Nixon C.,  2015, \mn@doi [ApJ] {10.1088/2041-8205/808/1/L11},
  808, L11

\bibitem[\protect\citeauthoryear{Debes, Hoard, Wachter, Leisawitz  \&
  Cohen}{Debes et~al.}{2011}]{Debes2011}
Debes J.~H.,  Hoard D.~W.,  Wachter S.,  Leisawitz D.~T.,   Cohen M.,  2011,
  \mn@doi [ApJS] {10.1088/0067-0049/197/2/38}, 197, 38

\bibitem[\protect\citeauthoryear{Dennihy, Debes, Dunlap, Dufour, Teske  \&
  Clemens}{Dennihy et~al.}{2016}]{Dennihy2016}
Dennihy E.,  Debes J.~H.,  Dunlap B.~H.,  Dufour P.,  Teske J.~K.,   Clemens
  J.~C.,  2016, \mn@doi [ApJ] {10.3847/0004-637X/831/1/31}, 831

\bibitem[\protect\citeauthoryear{Dennihy, Clemens, Debes, Dunlap, Kilkenny,
  O'Brien  \& Fuchs}{Dennihy et~al.}{2017}]{Dennihy2017}
Dennihy E.,  Clemens J.~C.,  Debes J.~H.,  Dunlap B.~H.,  Kilkenny D.,  O'Brien
  P.~C.,   Fuchs J.~T.,  2017, \mn@doi [ApJ] {10.3847/1538-4357/aa8ef2}, 849,
  77

\bibitem[\protect\citeauthoryear{Dennihy, Clemens, Dunlap, Fanale, Fuchs  \&
  Hermes}{Dennihy et~al.}{2018}]{Dennihy2018}
Dennihy E.,  Clemens J.~C.,  Dunlap B.~H.,  Fanale S.~M.,  Fuchs J.~T.,
  Hermes J.~J.,  2018, \mn@doi [ApJ] {10.3847/1538-4357/aaa89b}, 854, 40

\bibitem[\protect\citeauthoryear{Dennihy, Farihi, Pietro, Fusillo  \&
  Debes}{Dennihy et~al.}{2020}]{Dennihy2020}
Dennihy E.,  Farihi J.,  Pietro N.,  Fusillo G.,   Debes J.~H.,  2020, \mn@doi
  [ApJ] {10.3847/1538-4357/ab7249}, 891, 97

\bibitem[\protect\citeauthoryear{Drake et~al.,}{Drake et~al.}{2009}]{Drake2009}
Drake A.~J.,  et~al., 2009, \mn@doi [ApJ] {10.1088/0004-637X/696/1/870}, 696,
  870

\bibitem[\protect\citeauthoryear{Dufour, Kilic, Fontaine, Bergeron, Lachapelle,
  Kleinman  \& Leggett}{Dufour et~al.}{2010}]{Dufour2010}
Dufour P.,  Kilic M.,  Fontaine G.,  Bergeron P.,  Lachapelle F.-R.,  Kleinman
  S.~J.,   Leggett S.~K.,  2010, \mn@doi [ApJ] {10.1088/0004-637X/719/1/803},
  719, 803

\bibitem[\protect\citeauthoryear{Farihi}{Farihi}{2009}]{Farihi2009IRTF}
Farihi J.,  2009, \mn@doi [MNRAS] {10.1111/j.1365-2966.2009.15250.x}, 398, 2091

\bibitem[\protect\citeauthoryear{Farihi}{Farihi}{2016}]{Farihi2016}
Farihi J.,  2016, \mn@doi [New Astron. Rev.] {10.1016/j.newar.2016.03.001}, 71,
  9

\bibitem[\protect\citeauthoryear{Farihi, Becklin  \& Zuckerman}{Farihi
  et~al.}{2005}]{Farihi2005}
Farihi J.,  Becklin E.~E.,   Zuckerman B.,  2005, \mn@doi [ApJS]
  {10.1086/444362}, 161, 394

\bibitem[\protect\citeauthoryear{Farihi, Zuckerman  \& Becklin}{Farihi
  et~al.}{2008}]{Farihi2008}
Farihi J.,  Zuckerman B.,   Becklin E.~E.,  2008, \mn@doi [ApJ]
  {10.1086/521715}, 674, 431

\bibitem[\protect\citeauthoryear{Farihi, Jura  \& Zuckerman}{Farihi
  et~al.}{2009}]{Farihi2009Spitzer}
Farihi J.,  Jura M.,   Zuckerman B.,  2009, \mn@doi [ApJ]
  {10.1088/0004-637X/694/2/805}, 694, 805

\bibitem[\protect\citeauthoryear{Farihi, Jura, Lee  \& Zuckerman}{Farihi
  et~al.}{2010}]{Farihi2010}
Farihi J.,  Jura M.,  Lee J.-E.,   Zuckerman B.,  2010, \mn@doi [ApJ]
  {10.1088/0004-637X/714/2/1386}, 714, 1386

\bibitem[\protect\citeauthoryear{Farihi, Brinkworth, G{\"{a}}nsicke, Marsh,
  Girven, Hoard, Klein  \& Koester}{Farihi et~al.}{2011}]{Farihi2011GD61}
Farihi J.,  Brinkworth C.~S.,  G{\"{a}}nsicke B.~T.,  Marsh T.~R.,  Girven J.,
  Hoard D.~W.,  Klein B.,   Koester D.,  2011, \mn@doi [ApJ]
  {10.1088/2041-8205/728/1/L8}, 728, L8

\bibitem[\protect\citeauthoryear{Farihi, G{\"{a}}nsicke, Steele, Girven,
  Burleigh, Breedt  \& Koester}{Farihi et~al.}{2012}]{Farihi2012}
Farihi J.,  G{\"{a}}nsicke B.~T.,  Steele P.~R.,  Girven J.,  Burleigh M.~R.,
  Breedt E.,   Koester D.,  2012, \mn@doi [MNRAS]
  {10.1111/j.1365-2966.2012.20421.x}, 421, 1635

\bibitem[\protect\citeauthoryear{Farihi, Parsons  \& G{\"{a}}nsicke}{Farihi
  et~al.}{2017}]{Farihi2017SDSS1557}
Farihi J.,  Parsons S.~G.,   G{\"{a}}nsicke B.~T.,  2017, \mn@doi [Nature
  Astronomy] {10.1038/s41550-016-0032}, 1, 0032

\bibitem[\protect\citeauthoryear{Farihi et~al.,}{Farihi
  et~al.}{2018}]{Farihi2018GD56}
Farihi J.,  et~al., 2018, \mn@doi [MNRAS] {10.1093/mnras/sty2331}, 481, 2601

\bibitem[\protect\citeauthoryear{Fazio et~al.,}{Fazio et~al.}{2004}]{Fazio2004}
Fazio G.~G.,  et~al., 2004, \mn@doi [ApJS] {10.1086/422843}, 154, 10

\bibitem[\protect\citeauthoryear{Florez \& Wilson}{Florez \&
  Wilson}{2020}]{Florez2020}
Florez L.,  Wilson D.,  2020, in American Astronomical Society Meeting
  Abstracts. American Astronomical Society Meeting Abstracts.
p. 170.01

\bibitem[\protect\citeauthoryear{G{\"{a}}nsicke}{G{\"{a}}nsicke}{2011}]{Gansicke2011}
G{\"{a}}nsicke B.~T.,  2011, in Planetary Systems beyond the Main Sequence.
  AIP, pp 211--214, \mn@doi{10.1063/1.3556202}

\bibitem[\protect\citeauthoryear{G{\"{a}}nsicke, Marsh, Southworth  \&
  Rebassa-Mansergas}{G{\"{a}}nsicke et~al.}{2006}]{Gansicke2006}
G{\"{a}}nsicke B.~T.,  Marsh T.~R.,  Southworth J.,   Rebassa-Mansergas A.,
  2006, \mn@doi [Science] {10.1126/science.1135033}, 314, 1908

\bibitem[\protect\citeauthoryear{G{\"{a}}nsicke, Marsh  \&
  Southworth}{G{\"{a}}nsicke et~al.}{2007}]{Gansicke2007}
G{\"{a}}nsicke B.~T.,  Marsh T.~R.,   Southworth J.,  2007, \mn@doi [MNRAS]
  {10.1111/j.1745-3933.2007.00343.x}, 380, L35

\bibitem[\protect\citeauthoryear{G{\"{a}}nsicke, Koester, Marsh,
  Rebassa-Mansergas  \& Southworth}{G{\"{a}}nsicke et~al.}{2008}]{Gansicke2008}
G{\"{a}}nsicke B.~T.,  Koester D.,  Marsh T.~R.,  Rebassa-Mansergas A.,
  Southworth J.,  2008, \mn@doi [MNRAS] {10.1111/j.1745-3933.2008.00565.x},
  391, L103

\bibitem[\protect\citeauthoryear{G{\"{a}}nsicke, Koester, Farihi, Girven,
  Parsons  \& Breedt}{G{\"{a}}nsicke et~al.}{2012}]{Gansicke2012}
G{\"{a}}nsicke B.~T.,  Koester D.,  Farihi J.,  Girven J.,  Parsons S.~G.,
  Breedt E.,  2012, \mn@doi [MNRAS] {10.1111/j.1365-2966.2012.21201.x}, 424,
  333

\bibitem[\protect\citeauthoryear{Girven, G{\"{a}}nsicke, Steeghs  \&
  Koester}{Girven et~al.}{2011}]{Girven2011}
Girven J.,  G{\"{a}}nsicke B.~T.,  Steeghs D.,   Koester D.,  2011, \mn@doi
  [MNRAS] {10.1111/j.1365-2966.2011.19337.x}, 417, 1210

\bibitem[\protect\citeauthoryear{Girven, Brinkworth, Farihi, G{\"{a}}nsicke,
  Hoard, Marsh  \& Koester}{Girven et~al.}{2012}]{Girven2012}
Girven J.,  Brinkworth C.~S.,  Farihi J.,  G{\"{a}}nsicke B.~T.,  Hoard D.~W.,
  Marsh T.~R.,   Koester D.,  2012, \mn@doi [ApJ]
  {10.1088/0004-637X/749/2/154}, 749, 154

\bibitem[\protect\citeauthoryear{Guo, Tziamtzis, Wang, Liu, Zhao  \& Wang}{Guo
  et~al.}{2015}]{Guo2015}
Guo J.,  Tziamtzis A.,  Wang Z.,  Liu J.,  Zhao J.,   Wang S.,  2015, \mn@doi
  [ApJ] {10.1088/2041-8205/810/2/L17}, 810, L17

\bibitem[\protect\citeauthoryear{Hahn \& Rettig}{Hahn \&
  Rettig}{1998}]{Hahn1998}
Hahn J.~M.,  Rettig T.~W.,  1998, \mn@doi [Planet. Space Sci.]
  {10.1016/s0032-0633(98)00055-5}, 46, 1677

\bibitem[\protect\citeauthoryear{Hermes et~al.,}{Hermes
  et~al.}{2017}]{Hermes2017}
Hermes J.~J.,  et~al., 2017, \mn@doi [ApJS] {10.3847/1538-4365/aa8bb5}, 232, 23

\bibitem[\protect\citeauthoryear{Hoard, Debes, Wachter, Leisawitz  \&
  Cohen}{Hoard et~al.}{2013}]{Hoard2013}
Hoard D.~W.,  Debes J.~H.,  Wachter S.,  Leisawitz D.~T.,   Cohen M.,  2013,
  \mn@doi [ApJ] {10.1088/0004-637X/770/1/21}, 770, 21

\bibitem[\protect\citeauthoryear{Hollands, G{\"{a}}nsicke  \& Koester}{Hollands
  et~al.}{2018}]{Hollands2018analysis}
Hollands M.~A.,  G{\"{a}}nsicke B.~T.,   Koester D.,  2018, \mn@doi [MNRAS]
  {10.1093/mnras/sty592}, 477, 93

\bibitem[\protect\citeauthoryear{Honeycutt}{Honeycutt}{1992}]{Honeycutt1992}
Honeycutt R.~K.,  1992, \mn@doi [PASP] {10.1086/133015}, 104, 435

\bibitem[\protect\citeauthoryear{Jura}{Jura}{2003}]{Jura2003}
Jura M.,  2003, \mn@doi [ApJ] {10.1086/374036}, 584, L91

\bibitem[\protect\citeauthoryear{Jura \& Young}{Jura \&
  Young}{2014}]{Jura2014review}
Jura M.,  Young E.~D.,  2014, \mn@doi [Annu. Rev. Earth Planet. Sci.]
  {10.1146/annurev-earth-060313-054740}, 42, 45

\bibitem[\protect\citeauthoryear{Jura, Farihi, Zuckerman  \& Becklin}{Jura
  et~al.}{2007a}]{Jura2007GD362}
Jura M.,  Farihi J.,  Zuckerman B.,   Becklin E.~E.,  2007a, \mn@doi [AJ]
  {10.1086/512734}, 133, 1927

\bibitem[\protect\citeauthoryear{Jura, Farihi  \& Zuckerman}{Jura
  et~al.}{2007b}]{Jura2007Spitzer11stars}
Jura M.,  Farihi J.,   Zuckerman B.,  2007b, \mn@doi [ApJ] {10.1086/518767},
  663, 1285

\bibitem[\protect\citeauthoryear{Kenyon \& Bromley}{Kenyon \&
  Bromley}{2017a}]{Kenyon2017collisions}
Kenyon S.~J.,  Bromley B.~C.,  2017a, \mn@doi [ApJ] {10.3847/1538-4357/aa7b85},
  844, 116

\bibitem[\protect\citeauthoryear{Kenyon \& Bromley}{Kenyon \&
  Bromley}{2017b}]{Kenyon2017gas}
Kenyon S.~J.,  Bromley B.~C.,  2017b, \mn@doi [ApJ] {10.3847/1538-4357/aa9570},
  850, 50

\bibitem[\protect\citeauthoryear{Kilic \& Redfield}{Kilic \&
  Redfield}{2007}]{Kilic2007}
Kilic M.,  Redfield S.,  2007, \mn@doi [ApJ] {10.1086/513008}, 660, 641

\bibitem[\protect\citeauthoryear{Kilic, von Hippel, Leggett  \& Winget}{Kilic
  et~al.}{2005}]{Kilic2005}
Kilic M.,  von Hippel T.,  Leggett S.~K.,   Winget D.~E.,  2005, \mn@doi [ApJ]
  {10.1086/497825}, 632, L115

\bibitem[\protect\citeauthoryear{Kilic, {Von Hippel}, Leggett  \& Winget}{Kilic
  et~al.}{2006}]{Kilic2006DAZconnection}
Kilic M.,  {Von Hippel} T.,  Leggett S.~K.,   Winget D.~E.,  2006, ApJ, 646,
  474

\bibitem[\protect\citeauthoryear{Kilic, Patterson, Barber, Leggett  \&
  Dufour}{Kilic et~al.}{2012}]{Kilic2012}
Kilic M.,  Patterson A.~J.,  Barber S.,  Leggett S.~K.,   Dufour P.,  2012,
  \mn@doi [MNRAS] {10.1111/j.1745-3933.2011.01177.x}, 419, 59

\bibitem[\protect\citeauthoryear{Koester}{Koester}{2010}]{Koester2010}
Koester D.,  2010, Mem. Soc. Astron. Ital., 81, 921

\bibitem[\protect\citeauthoryear{Koz{\l}owski et~al.,}{Koz{\l}owski
  et~al.}{2010}]{Kozowski2010}
Koz{\l}owski S.,  et~al., 2010, \mn@doi [ApJ] {10.1088/0004-637X/716/1/530},
  716, 530

\bibitem[\protect\citeauthoryear{Law et~al.,}{Law et~al.}{2009}]{Law2009}
Law N.~M.,  et~al., 2009, \mn@doi [PASP] {10.1086/648598}, 121, 1395

\bibitem[\protect\citeauthoryear{Malamud \& Perets}{Malamud \&
  Perets}{2020a}]{Malamud2020A}
Malamud U.,  Perets H.~B.,  2020a, \mn@doi [MNRAS] {10.1093/mnras/staa142},
  492, 5561

\bibitem[\protect\citeauthoryear{Malamud \& Perets}{Malamud \&
  Perets}{2020b}]{Malamud2020B}
Malamud U.,  Perets H.~B.,  2020b, \mn@doi [MNRAS] {10.1093/mnras/staa143},
  493, 698

\bibitem[\protect\citeauthoryear{Manser et~al.,}{Manser
  et~al.}{2016}]{Manser2016a}
Manser C.~J.,  et~al., 2016, \mn@doi [MNRAS] {10.1093/mnras/stv2603}, 455, 4467

\bibitem[\protect\citeauthoryear{Manser et~al.,}{Manser
  et~al.}{2019}]{Manser2019}
Manser C.~J.,  et~al., 2019, \mn@doi [Science (New York, N.Y.)]
  {10.1126/science.aat5330}, 364, 66

\bibitem[\protect\citeauthoryear{Manser, G{\"{a}}nsicke, {Gentile Fusillo},
  Ashley, Breedt, Hollands, Izquierdo  \& Pelisoli}{Manser
  et~al.}{2020}]{Manser2020}
Manser C.~J.,  G{\"{a}}nsicke B.~T.,  {Gentile Fusillo} N.~P.,  Ashley R.,
  Breedt E.,  Hollands M.,  Izquierdo P.,   Pelisoli I.,  2020, \mn@doi [MNRAS]
  {10.1093/mnras/staa359}, 493, 2127

\bibitem[\protect\citeauthoryear{Masci et~al.,}{Masci et~al.}{2019}]{Masci2018}
Masci F.~J.,  et~al., 2019, \mn@doi [PASP] {10.1088/1538-3873/aae8ac}, 131

\bibitem[\protect\citeauthoryear{Melis, Jura, Albert, Klein  \&
  Zuckerman}{Melis et~al.}{2010}]{Melis2010}
Melis C.,  Jura M.,  Albert L.,  Klein B.,   Zuckerman B.,  2010, \mn@doi [ApJ]
  {10.1088/0004-637X/722/2/1078}, 722, 1078

\bibitem[\protect\citeauthoryear{Melis, Farihi, Dufour, Zuckerman, Burgasser,
  Bergeron, Bochanski  \& Simcoe}{Melis et~al.}{2011}]{Melis2011}
Melis C.,  Farihi J.,  Dufour P.,  Zuckerman B.,  Burgasser A.~J.,  Bergeron
  P.,  Bochanski J.,   Simcoe R.,  2011, \mn@doi [ApJ]
  {10.1088/0004-637X/732/2/90}, 732, 90

\bibitem[\protect\citeauthoryear{Melis et~al.,}{Melis et~al.}{2012}]{Melis2012}
Melis C.,  et~al., 2012, \mn@doi [ApJ] {10.1088/2041-8205/751/1/L4}, 751, L4

\bibitem[\protect\citeauthoryear{Metzger, Rafikov  \& Bochkarev}{Metzger
  et~al.}{2012}]{Metzger2012}
Metzger B.~D.,  Rafikov R.~R.,   Bochkarev K.~V.,  2012, \mn@doi [MNRAS]
  {10.1111/j.1365-2966.2012.20895.x}, 423, 505

\bibitem[\protect\citeauthoryear{Miranda \& Rafikov}{Miranda \&
  Rafikov}{2018}]{Miranda2018}
Miranda R.,  Rafikov R.~R.,  2018, \mn@doi [ApJ] {10.3847/1538-4357/aab9a2},
  857, 135

\bibitem[\protect\citeauthoryear{Mullally, Kilic, Reach, Kuchner, von Hippel,
  Burrows  \& Winget}{Mullally et~al.}{2007}]{Mullally2007}
Mullally F.,  Kilic M.,  Reach W.~T.,  Kuchner M.~J.,  von Hippel T.,  Burrows
  A.,   Winget D.~E.,  2007, \mn@doi [ApJS] {10.1086/511858}, 171, 206

\bibitem[\protect\citeauthoryear{Nixon, Pringle, Coughlin, Swan  \&
  Farihi}{Nixon et~al.}{2020}]{Nixon2020}
Nixon C.~J.,  Pringle J.~E.,  Coughlin E.~R.,  Swan A.,   Farihi J.,  2020,
  {preprint} (\mn@eprint {arXiv} {2006.07639})

\bibitem[\protect\citeauthoryear{Patten et~al.,}{Patten
  et~al.}{2006}]{Patten2006}
Patten B.~M.,  et~al., 2006, \mn@doi [ApJ] {10.1086/507264}, 651, 502

\bibitem[\protect\citeauthoryear{Patterson, Zuckerman, Becklin, Tholen  \&
  Hawarden}{Patterson et~al.}{1991}]{Patterson1991}
Patterson J.,  Zuckerman B.,  Becklin E.~E.,  Tholen D.~J.,   Hawarden T.,
  1991, \mn@doi [ApJ] {10.1086/170122}, 374, 330

\bibitem[\protect\citeauthoryear{Polimera, Sarajedini, Ashby, Willner  \&
  Fazio}{Polimera et~al.}{2018}]{Polimera2018}
Polimera M.,  Sarajedini V.,  Ashby M. L.~N.,  Willner S.~P.,   Fazio G.~G.,
  2018, \mn@doi [MNRAS] {10.1093/mnras/sty164}, 476, 1111

\bibitem[\protect\citeauthoryear{Rafikov}{Rafikov}{2011}]{Rafikov2011runaway}
Rafikov R.~R.,  2011, \mn@doi [MNRAS] {10.1111/j.1745-3933.2011.01096.x}, 416,
  L55

\bibitem[\protect\citeauthoryear{Reach et~al.,}{Reach
  et~al.}{2005}]{Reach2005IRAC}
Reach W.~T.,  et~al., 2005, \mn@doi [PASP] {10.1086/432670}, 117, 978

\bibitem[\protect\citeauthoryear{Rebassa-Mansergas, Solano, Xu, Rodrigo,
  Jim{\'{e}}nez-Esteban  \& Torres}{Rebassa-Mansergas
  et~al.}{2019}]{Rebassa-Mansergas2019}
Rebassa-Mansergas A.,  Solano E.,  Xu S.,  Rodrigo C.,  Jim{\'{e}}nez-Esteban
  F.~M.,   Torres S.,  2019, \mn@doi [MNRAS] {10.1093/mnras/stz2423}, 489, 3990

\bibitem[\protect\citeauthoryear{Ricker et~al.,}{Ricker
  et~al.}{2014}]{Ricker2014}
Ricker G.~R.,  et~al., 2014, \mn@doi [Journal of Astronomical Telescopes,
  Instruments, and Systems] {10.1117/1.jatis.1.1.014003}, 1, 014003

\bibitem[\protect\citeauthoryear{Rocchetto, Farihi, Gansicke, Bergfors,
  G{\"{a}}nsicke  \& Bergfors}{Rocchetto et~al.}{2015}]{Rocchetto2015}
Rocchetto M.,  Farihi J.,  Gansicke B.~T.,  Bergfors C.,  G{\"{a}}nsicke B.~T.,
    Bergfors C.,  2015, \mn@doi [MNRAS] {10.1093/mnras/stv282}, 449, 574

\bibitem[\protect\citeauthoryear{Rogers, Xu, Bonsor, Hodgkin, Su, von Hippel
  \& Jura}{Rogers et~al.}{2020}]{Rogers2020}
Rogers L.~K.,  Xu S.,  Bonsor A.,  Hodgkin S.,  Su K. Y.~L.,  von Hippel T.,
  Jura M.,  2020, \mn@doi [MNRAS] {10.1093/mnras/staa873}, 494, 2861

\bibitem[\protect\citeauthoryear{Swan, Farihi  \& Wilson}{Swan
  et~al.}{2019}]{Swan2019}
Swan A.,  Farihi J.,   Wilson T.~G.,  2019, \mn@doi [MNRAS]
  {10.1093/mnrasl/slz014}, 484, L109

\bibitem[\protect\citeauthoryear{Tielens, McKee, Seab  \& Hollenbach}{Tielens
  et~al.}{1994}]{Tielens1994}
Tielens A. G. G.~M.,  McKee C.~F.,  Seab C.~G.,   Hollenbach D.~J.,  1994,
  \mn@doi [ApJ] {10.1086/174488}, 431, 321

\bibitem[\protect\citeauthoryear{Tokunaga, Becklin  \& Zuckerman}{Tokunaga
  et~al.}{1990}]{Tokunaga1990}
Tokunaga A.~T.,  Becklin E.~E.,   Zuckerman B.,  1990, \mn@doi [ApJ]
  {10.1086/185770}, 358, L21

\bibitem[\protect\citeauthoryear{Vanderburg et~al.,}{Vanderburg
  et~al.}{2015}]{Vanderburg2015}
Vanderburg A.,  et~al., 2015, \mn@doi [Nature] {10.1038/nature15527}, 526, 546

\bibitem[\protect\citeauthoryear{Wang et~al.,}{Wang
  et~al.}{2019}]{Wang2019_0145}
Wang T.-G.,  et~al., 2019, \mn@doi [ApJ] {10.3847/2041-8213/ab53ed}, 886, L5

\bibitem[\protect\citeauthoryear{Werner et~al.,}{Werner
  et~al.}{2004}]{Werner2004Spitzer}
Werner M.~W.,  et~al., 2004, \mn@doi [ApJS] {10.1086/422992}, 154, 1

\bibitem[\protect\citeauthoryear{Wickramasinghe, Hoyle  \&
  Al-Mufti}{Wickramasinghe et~al.}{1988}]{Wickramasinghe1988}
Wickramasinghe N.~C.,  Hoyle F.,   Al-Mufti S.,  1988, \mn@doi [Astrophysics
  and Space Science] {10.1007/BF00636767}, 143, 193

\bibitem[\protect\citeauthoryear{Wilson, G{\"{a}}nsicke, Koester, Raddi,
  Breedt, Southworth  \& Parsons}{Wilson et~al.}{2014}]{Wilson2014}
Wilson D.~J.,  G{\"{a}}nsicke B.~T.,  Koester D.,  Raddi R.,  Breedt E.,
  Southworth J.,   Parsons S.~G.,  2014, \mn@doi [MNRAS]
  {10.1093/mnras/stu1876}, 445, 1878

\bibitem[\protect\citeauthoryear{Wilson, Farihi, G{\"{a}}nsicke  \&
  Swan}{Wilson et~al.}{2019}]{Wilson2019}
Wilson T.~G.,  Farihi J.,  G{\"{a}}nsicke B.~T.,   Swan A.,  2019, \mn@doi
  [MNRAS] {10.1093/mnras/stz1050}, 487, 133

\bibitem[\protect\citeauthoryear{Wyatt}{Wyatt}{2008}]{Wyatt2008}
Wyatt M.~C.,  2008, \mn@doi [ARA{\&}A]
  {10.1146/annurev.astro.45.051806.110525}, 46, 339

\bibitem[\protect\citeauthoryear{Xu \& Jura}{Xu \& Jura}{2012}]{Xu2012}
Xu S.,  Jura M.,  2012, \mn@doi [ApJ] {10.1088/0004-637X/745/1/88}, 745, 88

\bibitem[\protect\citeauthoryear{Xu \& Jura}{Xu \& Jura}{2014}]{Xu2014drop}
Xu S.,  Jura M.,  2014, \mn@doi [ApJ] {10.1088/2041-8205/792/2/L39}, 792, L39

\bibitem[\protect\citeauthoryear{Xu et~al.,}{Xu et~al.}{2018a}]{Xu2018WD1145}
Xu S.,  et~al., 2018a, \mn@doi [MNRAS] {10.1093/mnras/stx3023}, 474, 4795

\bibitem[\protect\citeauthoryear{Xu et~al.,}{Xu
  et~al.}{2018b}]{Xu2018IRvariability}
Xu S.,  et~al., 2018b, \mn@doi [ApJ] {10.3847/1538-4357/aadcfe}, 866, 108

\bibitem[\protect\citeauthoryear{Xu et~al.,}{Xu
  et~al.}{2019a}]{Xu2019UVtransits1145}
Xu S.,  et~al., 2019a, \mn@doi [AJ] {10.3847/1538-3881/ab1b36}, 157, 255

\bibitem[\protect\citeauthoryear{Xu, Dufour, Klein, Melis, Monson, Zuckerman,
  Young  \& Jura}{Xu et~al.}{2019b}]{Xu2019composition}
Xu S.,  Dufour P.,  Klein B.,  Melis C.,  Monson N.~N.,  Zuckerman B.,  Young
  E.~D.,   Jura M.~A.,  2019b, \mn@doi [AJ] {10.3847/1538-3881/ab4cee}, 158,
  242

\bibitem[\protect\citeauthoryear{Zuckerman \& Becklin}{Zuckerman \&
  Becklin}{1987}]{Zuckerman1987}
Zuckerman B.,  Becklin E.~E.,  1987, \mn@doi [Nature] {10.1038/330138a0}, 330,
  138

\makeatother
\end{thebibliography}




\appendix


\begin{table*}
\caption{Target list for program~13216, number of observations, and maximum flux changes observed.}
\resizebox{\textwidth}{!}{ 
\begin{tabular}{llccSScSSl}
\hline
&&&\multicolumn{3}{c}{............................ {3.6\,\micron} ............................}&\multicolumn{3}{c}{............................ {4.5\,\micron} ............................}\\
WD         & Other name  & Type$^{\rm a}$ & Epochs & {Change (\%)}   & {Significance ($\upsigma$)} & Epochs & {Change (\%)} & {Significance ($\upsigma$)} & References\\
\hline
0106$-$328 &                             &      &      2 &               9 &                       6 &      3 &              10 &                       6 & 1 \\
0110$-$565 &                             &      &      3 &              15 &                       9 &      3 &              20 &                      12 & 2 \\
0146+187   &                      GD\,16 &      &      2 &             0.3 &                     0.2 &      3 &               7 &                       4 & 1 \\
0246+734   &                             &      &      2 &               6 &                       3 &      3 &              17 &                       5 & 3 \\
0300$-$013 &                      GD\,40 &      &      2 &               9 &                       6 &      2 &              13 &                       8 & 4 \\
0307+078   &               HS\,0307+0746 &      &      2 &               3 &                       2 &      2 &               9 &                       5 & 5 \\
0408$-$041 &                      GD\,56 &      &      8 &              16 &                       9 &      9 &              26 &                      13 & 6 \\
0420+520   &                             &      &      2 &               3 &                       2 &      1 &                 &                         & 7 \\
0420$-$731 &                             &      &      1 &                 &                         &      1 &                 &                         & 7 \\
0435+410   &                      GD\,61 &      &      2 &               1 &                     0.8 &      3 &               3 &                       2 & 8 \\
0536$-$479 &            EC\,05365$-$4749 &      &      1 &                 &                         &      1 &                 &                         & 9 \\
0735+187   &   SDSS\,J073842.56+183509.6 &    E &      3 &              40 &                      21 &      3 &              46 &                      25 & 10, 11 \\
0842+231   &                    Ton\,345 &    E &      1 &                 &                         &      2 &              12 &                       7 & 12, 13 \\
0842+572   &                             &    E &      1 &                 &                         &      1 &                 &                         & 14 \\
0843+516   &                             &      &      2 &             0.2 &                     0.1 &      2 &               7 &                       5 & 15 \\
0956$-$017 & SDSS\,J095904.69$-$020047.6 &    E &      3 &              55 &                      22 &      3 &              50 &                      24 & 16, 17\\
1015+161   &                             &      &      2 &               5 &                       3 &      3 &               7 &                       5 & 4 \\
1018+410   &                             &      &      3 &               8 &                       4 &      3 &               9 &                       6 & 18 \\
1041+091   &   SDSS\,J104341.53+085558.2 &    E &      2 &              10 &                       3 &      2 &               4 &                       2 & 19, 20 \\
1116+026   &                      GD 133 &    V &      2 &               2 &                       1 &      3 &             0.6 &                     0.4 & 4 \\
1141+057$^{\rm b}$ & SDSS\,J114404.74+052951.6 & E &   1 &                 &                         &      1 &                 &                         & 21 \\
1145+017   &                             &      &      1 &                 &                         &      7 &               7 &                       3 & 22 \\
1145+288   &                             &    V &      1 &                 &                         &      1 &                 &                         & 23 \\
1150$-$153 &                             &    V &      5 &               7 &                       4 &      5 &               7 &                       4 & 24 \\
1219+130   &   SDSS\,J122150.81+124513.3 &      &      2 &               2 &                       1 &      2 &               4 &                       3 & 17 \\
1225$-$079 &                             &      &      2 &               1 &                     0.7 &      3 &               5 &                       3 & 5 \\
1226+110   &                             &    E &      3 &              26 &                      13 &      3 &              21 &                      13 & 25, 26\\
1232+563   &                             &      &      2 &               4 &                       2 &      2 &               8 &                       5 & 27 \\
1349$-$230 &                             &    E &      1 &                 &                         &      2 &              11 &                       7 & 2, 28 \\
1455+298   &                     G166-58 &      &      2 &               2 &                       2 &      3 &               3 &                       2 & 29 \\
1457$-$086 &                             &      &      2 &               2 &                       1 &      2 &             0.8 &                     0.5 & 1 \\
1536+520   &                             &      &      1 &                 &                         &      1 &                 &                         & 27 \\
1541+651   &                PG\,1541+651 &    V &      2 &               8 &                       5 &      2 &              11 &                       7 & 30 \\
1551+175   &                             &      &      2 &             0.2 &                    0.03 &      2 &               8 &                       2 & 3 \\
1554+094   &                             &      &      4 &              56 &                       9 &      4 &              44 &                      14 & 17 \\
1615+164   &   SDSS\,J161717.04+162022.4 &    E &      2 &              12 &                       7 &      2 &               8 &                       5 & 12, 31 \\
1729+371   &                     GD\,362 &      &      6 &               5 &                       3 &      8 &              11 &                       6 & 32, 33 \\
1929+012   &     GALEX\,J193156.8+011745 &      &      2 &               1 &                     0.7 &      2 &               2 &                       2 & 34 \\
2115$-$560 &                             &      &      6 &               2 &                       2 &      7 &               7 &                       4 & 35 \\
2132+096   &                             &      &      2 &               1 &                     0.9 &      2 &               1 &                     0.8 & 3 \\
2207+121   &                             &      &      2 &               3 &                       2 &      2 &             0.2 &                     0.1 & 15 \\
2221$-$165 &                             &      &      2 &               8 &                       6 &      2 &              12 &                       8 & 5 \\
2326+049   &                      G29-38 &    V &      7 &              12 &                       7 &      8 &               9 &                       4 & 36 \\
2329+407   &                             &      &      1 &                 &                         &      1 &                 &                         & 7 \\
\hline
\end{tabular}} 
\flushleft
\footnotesize {$^{\rm a}$Type~E stars show \ion{Ca}{ii}~emission. Type~V stars are known or suspected pulsators.

$^{\rm b}$Infared excess and \ion{Ca}{ii} emission are likely due to a low-mass companion rather than circumstellar debris; see Section~\ref{subsectionIndividualObjects}.
}

\smallskip
References:
(1)~\citep{Farihi2009IRTF};
(2)~\citep{Girven2012};
(3)~\citep{Bergfors2014};
(4)~\citep{Jura2007Spitzer11stars};
(5)~\citep{Farihi2010};
(6)~\citep{Kilic2006DAZconnection};
(7)~\citep{Hoard2013};
(8)~\citep{Farihi2011GD61};
(9)~\citep{Dennihy2016};
(10)~\citep{Dufour2010};
(11)~\citep{Gansicke2011};
(12)~\citep{Brinkworth2012};
(13)~\citep{Gansicke2008};
(14)~Boris Gansicke priv. comm.;
(15)~\citep{Xu2012};
(16)~\citep{Girven2011};
(17)~\citep{Farihi2012};
(18)~\citep{Rocchetto2015};
(19)~\citep{Melis2010};
(20)~\citep{Gansicke2007};
(21)~\citep{Guo2015};
(22)~\citep{Vanderburg2015};
(23)~\citep{Barber2014};
(24)~\citep{Kilic2007};
(25)~\citep{Brinkworth2009};
(26)~\citep{Gansicke2006};
(27)~\citep{Debes2011};
(28)~\citep{Melis2012};
(29)~\citep{Farihi2008};
(30)~\citep{Kilic2012};
(31)~\citep{Wilson2014};
(32)~\citep{Becklin2005};
(33)~\citep{Kilic2005};
(34)~\citep{Melis2011};
(35)~\citep{Mullally2007};
(36)~\citep{Zuckerman1987};

\label{tableResults}
\end{table*}


\bsp	
\label{lastpage}
\end{document}